\documentstyle[aps,manuscript,epsf]{revtex} 

\begin{document} 
\draft

\title{Attraction between DNA molecules mediated by multivalent ions} 

\author{E.Allahyarov $^{1}$,\,\, G.Gompper $^{1}$,\,\, H.L\"owen $^{2}$}
\address{{1} Institut\, f\"ur\, Festk\"orperforschung, Forschungszentrum 
J\"ulich, \,\mbox{D-52425} \, J\"ulich, Germany }
\address{{2} Institut\, f\"{u}r\, Theoretische \,Physik
  II,\,Heinrich-Heine-Universit\"{a}t\,
 D\"{u}sseldorf,\,\mbox{D-40225}\,D\"{u}sseldorf, \,Germany}

\date{\today}   

\maketitle
\begin{abstract}
The effective force between two parallel DNA molecules
is calculated as a function of their mutual separation for
different valencies of counter- and salt ions and 
different salt concentrations. Computer simulations of the
primitive model are used and the shape of the DNA molecules 
is accurately modelled using different geometrical shapes.
We find that multivalent ions induce a significant attraction
between the DNA molecules whose strength can be tuned
by the averaged valency of the ions. The  physical origin 
of the attraction is traced back either to
electrostatics or to entropic contributions. For multivalent counter-
and monovalent salt ions, we find a salt-induced stabilization effect: 
 the force is first attractive
but gets repulsive for increasing salt concentration. Furthermore, we show
that the multivalent-ion-induced
 attraction does not necessarily correlate with DNA overcharging.
\end{abstract}

\section{Introduction}
During the last decade the question that concerns the possible existence
of long-ranged attractive interactions between similarly charged objects in
electrolyte solutions has been intensely debated. Experimental evidence
of such an attraction is seen for deoxyribose nucleic acid (DNA) molecules
\cite{pelta1996,bloomfield1991biopolymers,bloomfield1997,bloomfield1996cosb,shui1,tajmir},
colloidal rods \cite{wierenga1996}, charged clay particles
\cite{kjellander1988colloidintsci}, charged microspheres
\cite{kepler,crocer1996prl,larsen1997nature}
and charged plates \cite{wennerstrom1991acis,kekicheff}. 
In particular, DNA molecules in solution are a paradigm for 
 negatively charged polyelectrolytes due to
 ionization of its acidic phosphate groups \cite{saenger}.
 The DNA conformations display a considerable sensitivity to the ionic
surrounding. The mutual repulsion of DNA polyions has to be overcome
to form compact or condensed DNA bundles. 
Experiments show that DNA condensation
occurs when about 90 percent of its charge is neutralized by
condensed counterions \cite{pelta1996,bloomfield1997,benbasat1984,deng1999}. 
Such a strong neutralization of the DNA
charge could be achieved by divalent and higher-valent
counterions \cite{Solis1999,lamm1993}.
Besides of the phosphate neutralization, the multivalent
ions induce an additional attraction between the DNA macroions
 mediated by strong correlation effects
\cite{deng1999,nilsson,lyubartsev1995,Deng,khan,ha1998attraction,deserno2002attraction,guldbrand,bolhius}. 
Thus the small ions play a complex role in DNA-DNA interactions and
are not simply agents to screen the long-range
electrostatic interaction. For example, they adsorb onto the DNA surface and can
create bridges between the DNA molecules at small DNA-DNA separations,
resulting in an ion cross-link attraction \cite{deserno2002attraction,cai2002}.

The electrostatic interaction between highly charged
polyelectrolytes is usually
treated within the framework of classical double-layer theory
\cite{vervey}. 
This theory is based on the mean-field
Poisson-Boltzmann equation
\cite{reiner1993acis,mills,marcus1955,fogo,lin,gilson,wagner,vlachy1999}
and predicts a
repulsion between similarly charged macromolecules.
Though different modifications of Poisson-Boltzmann theory 
have been developed to account for ion-ion hard core correlations
\cite{gavryushov,rescic1997,das1995modifiedPB,Netz1},
an attractive contribution in the double-layer theory 
is usually introduced via the van der Waals 
interaction forces \cite{kepler,sood1991,coen1995,vlachy1993}.
 However, the van der Waals forces alone cannot explain the
experimentally observed attraction, since  
the Hamaker constant extracted from the
 experiments is artificially high
\cite{kepler,farnum1999bj,spalla2000hamaker,weiss2000cps}.

Theoretical investigations and numerical simulations indicate that an
attraction between similarly charged objects 
emerges beyond the mean-field approaches.
It is now a well
established fact that the charge correlations and fluctuations in
highly charged electrolytes can induce an attraction between the macroions 
\cite{ha1998attraction,hnc_attraction,pincus,harau2002jcp,kjellander1986,attard1988,podgornik1990,stevens1990,lau2000prl,roij2000,roij1997and1999,Spalla,hribar1997,netz1999epland1999preand2001epje,goulding1999epl,ariel2003,burak2003,gronbech,levin,delrow,Linse1999,kjellander1992chempot,rouzina,lukatsky1999fluctuation}.
Due to the very resemblance of the short DNA fragments to charged
rods, the latter is a widely used toy model for the DNA molecule in
theoretical treatments and computer simulations
\cite{deserno2002attraction,mills,gronbech,murthy,vlachy,paulsen1988,mills1986,paulsen1987,katchalsky}. 
However, the details of DNA, such as the discreteness and helical
structure of the DNA phosphate charges and the grooved shape of the DNA molecule,
become essential as one 
approaches its surface. In this case, strictly speaking, all atom DNA simulations
in molecular water would be a proper choice \cite{jaya}. Unfortunately
such sophisticated simulations can only be applied 
to small systems and small salt concentrations \cite{yang1995bj}.
Thus, first, a devision of a ``sophisticated'' DNA model, which goes beyond the simple
homogeneously charged cylinder model,
and second, an investigation of the interaction forces between such
DNA molecules, remains a challenging task. 

This paper is an extension of our previous works
 on DNA electrostatics \cite{ourfirstDNApaper,oursecondDNApaper}. In
 Ref.~\cite{ourfirstDNApaper}, 
 simulation results for the DNA-DNA interaction were compared with
 the predictions of different linear
 theories for the case of {\it monovalent} ions. It was shown that the DNA-DNA
 interaction, at separation distances smaller than the Debye screening
 length, differs from the predictions of mean-field theories. 
 This provides evidence that the intermolecular
 interaction depends not only on how many ions are in the DNA
 proximity (which is exactly what the ordinary linear theories rely on), 
but also on where those ions are located relative to the DNA
 structure, i.e., whether they 
penetrate into the grooves or not. In Ref.~\cite{oursecondDNApaper}, on
 the other hand, a detailed distribution of ions of
 different valencies and molarities near the DNA surface was explored
 for {\it a more realistic, grooved shape of the DNA molecule}.
The results obtained   
 indicate that the paths of
counterion and coion condensations  strongly
depend on the DNA  surface geometry. Taking this into account we expect
 that the implemented DNA models with different geometries will also
 affect significantly the effective DNA-DNA interaction. 
Thus, in this paper, we will focus on the
 mechanism of  attraction
between two DNA molecules shaped similar to models introduced in
Ref.~\cite{oursecondDNApaper}.
Our goal is to see the effects, which increasing detail of various DNA
 models have on the DNA-DNA interaction.
We show that the DNA shape is an essential
contributor to the interaction force for multivalent counterions,
whereas it has a minor effect on
the interaction force for added  multivalent salt.
The origin of the attraction in the simple and sophisticated DNA
models is different. For instance, a Coulomb depletion-like
attraction \cite{allah} for the salt-free case
depends on the implemented DNA model. It has been revealed that there is a
non-monotonic force-salt dependence at a fixed DNA-DNA separation for
added monovalent salt and divalent counterions. 
This is exemplified by the variation of the
interaction force from a strong attraction towards a strong repulsion
and a following decrease in magnitude. Detailed investigations connect
this ``salt-induced stabilization'' to the entropic part of the total
 interaction force.   
 We also address the competition between the  multivalent counterion and
 the multivalent salt-induced attractions. It is shown that the increase
 of the divalent salt concentration at a fixed monovalent ion
number drives the DNA-DNA interaction 
force into an attraction through the overcharging of DNA
 molecules. However, the DNA-DNA attraction induced by trivalent
 counterions decreases, while the DNA molecule gradually gets
 overcharged due to added divalent salt. 
 
The reminder of paper is organized as follows. 
 We give a short general overview of ion binding and DNA
 condensation in Section II.
The system parameters and quantities studied in the present work are
 discussed in Section III. 
Sections IV and V contain simulation details and the implemented
 simulation techniques. The specific DNA configurations
 at short DNA-DNA separations are discussed in Section VI.
Sections VII and VIII are devoted to simulation  
 results for monovalent and divalent salt ions respectively.
 We conclude in Section IX.

\section{ion binding and DNA condensation} 
There are essentially two contenders for the dominant attractive force
in the DNA condensation: hydration forces \cite{parsegian,schneider1998}
and correlated counterion charge
fluctuations. Throughout this paper we neglect the granular nature of
water and the solvent-induced forces \cite{gonzales-mozuelos2000},
 and concentrate only on the electrostatics of the DNA
condensation. The water dielectric
effects and hydration forces will be briefly (and qualitatively) discussed in Section IX.

Under physiological conditions, the DNA molecule is surrounded by an ionic
atmosphere with a Debye screening length $\lambda_D$
in the range of 
5\AA-10\AA. Within the distances $r<\lambda_D$ above  the DNA surface,
a nonlinear screening of the DNA phosphate charges takes
place. Hence, if the surface-to-surface separation between two DNA molecules
is less than $\lambda_D$, a nonlinear theory
\cite{barrat1996,oosawa1971,bloomfield1999enzymol,hecht1995} has to be applied.
 At surface-to-surface separation distances on the order of 
or beyond $\lambda_D$, Debye-H\"{u}ckel theory (based on a linearized
Poisson-Boltzmann treatment) is a reasonable
approximation to describe the ionic atmosphere around the DNA
molecule. 
In Ref.~\cite{ourfirstDNApaper} we have examined several mean-field
theories for their ability to match the numerically calculated DNA-DNA
interaction forces: the homogeneously
charged rod model, the Yukawa segment (YS) model, and the Kornyshev-Leikin
(KL) theory \cite{kor1}. For the case of an overall
monovalency of counterions and
salt ions, both the simulations \cite{ourfirstDNApaper} and the above
mentioned theories 
reveal repulsive forces between the DNA molecules for all
 mutual orientations and separation distances. We have shown that, except for
short separation distances, there is a qualitative agreement between the theoretical
and numerical results if a proper charge and size renormalization in
the former is performed.

For multivalent counterions and added salt ions, there is experimental
evidence that the DNA molecules attract each other.
Such an attraction is
completely missed in the linear theories such as the homogeneously charged
cylinder and YS model. In these theories
all the nonlinear salt effects are again accounted for through the phosphate charge and
screening length renormalization procedure. Only the mean-field KL
theory predicts a DNA-DNA attraction for some DNA-DNA separations and
 azimuthal molecular orientations.
In detail, the KL theory distinguishes between
strongly condensed (also called as bound or adsorbed) and 
a cloud of diffusive (non-bonded) counterions. A tight adsorption is
assumed to take place within the Stern
layer of thickness $\xi=A/4\pi \lambda_B$=2\AA, 
where $\lambda_B=e^2/(\epsilon k_B T)$  is the Bjerrum length, $A$ is an
average area per elementary charge on the DNA surface, $\epsilon$ is
a dielectric constant of solution and $k_BT$ is the thermal energy. 
The KL theory \cite{kor1} predicts an attractive force between the two
DNA molecules if the following conditions are fulfilled: 
i) more in-groove than on-strand
condensation, ii) the right complementary alignment of the
positively charged grooves on one helix facing the negatively charged
strand on the other helix. In other words, the KL theory assumes that
it is the DNA charge helicity that entails an intermolecular attraction 
for surface-to-surface distances in the range of $8-15$\AA. Theoretical results and
computer simulations 
\cite{wagner,ourfirstDNApaper,lyubar,montoro1998,gulbrand1989,jaya2,conrad1988biopol,hochberg},
however, indicate that  no charge-helicity effects extend further than few \AA \,
from the DNA surface. There is also experimental evidence
\cite{zakharova} that at surface-to-surface separation
distances comparable with
the Debye screening length, the DNA-DNA separation does not affect
the DNA orientation.

 The discreteness of the DNA phosphates,
explicitly taken into account by our DNA models, enhances the
counterion concentration \cite{lukatsky2002epl} and the surface
adsorption of ions \cite{kjellander2001solvent} through the increased
Coulomb coupling between the phosphates and the counterions. This boosts the 
counterion correlations 
near the DNA surface \cite{messina1}. Experiments indicate that the divalent
counterions, depending on their in-groove or on-strand localization
\cite{sponer2002}, have different impact on the DNA systems. Thus, the  
 transition metals with higher affinity to the
 DNA bases\cite{duguid1993,duguid1995}
 condense on DNA 
\cite{knoll1988book,rau1992biophys}, while alkali metals do
not \cite{bloomfield1996cosb,rau1992biophys}. 
On the other hand, the chemical identity of the cation
is a factor of minor importance compared with
the magnitude of their charge when $q_c>2$
\cite{bloomfield1997,bloomfield1996cosb,deng1999,widom,plum,braunlin1986biopolymers,arscott1995biopolymers}.
Thus the spermidine Spd$_{3+}$ and
spermine Spm$_{4+}$ ions, abundant in living cells
\cite{cohen1998book,tabor1984,marx1982,saminathan}, neutralize the negatively
charged DNA backbone predominantly via the non-specific (Coulomb) interaction
\cite{gosule1976and1978,wilson1,deng2000}.
 This is supported by new experiments \cite{Deng,raspaud1999}, polyelectrolyte
and counterion condensation theories
\cite{anderson1982arpc,manning1,manning1992,rouzina2}, and 
computer simulations \cite{lyubar,korolev}.

\section{system parameters}
\subsection{DNA Models.}
The B form of DNA has an inner core of radius 9\AA \, formed by
 nucleotide pairs, and two
sugar-phosphate strands spiralling around it. The latter form the well-known
double helix with a pitch length $P$ about 34\AA \, \cite{saenger}. There are two
 phosphate groups per base pair, and 10 base pairs per 
pitch length, or helical turn. The axial rise per base pair in the DNA
long axis is 3.4\AA,\, thus there is  one elementary charge per each
1.7\AA \, \cite{andrey}. The average value of the angle between the adjacent
base pairs is $36^0$ and the average distance between the neighboring
charges on the DNA surface is about 7\AA. This distance is much smaller than
the helical pitch and of the order of Debye screening length under
the physiological conditions. 
There is a small shift in the $z$ coordinate of two opposing
 phosphates belonging to different helices of DNA, $\delta
 z=0.34$\AA. 

Three DNA models, a 
cylinder model (CM), an extended cylinder model (ECM) and
the Montoro-Abascal model (MAM), are considered.
Our aim is to obtain a detailed understanding of the physical
mechanism of ion-mediated DNA interactions, in particular how the
geometry of different DNA models gives rise to new effects. 
The CM has a hard cylindrical core 
 of diameter $D=20$\AA \, and two strings of monovalent 
phosphates of size $d_p=0.4$\AA. The KL theory, and almost all the Poisson-Boltzmann
like theories and most of primitive model (PM) computer simulations, have utilized the
CM as a simple DNA model.  
 In the ECM, first designed by
Lyubartsev {\it et. al.} \cite{lyubar}, the helical grooves of DNA are 
incorporated through the shrinking of the DNA core to the size $D=17.8$\AA \, and
swelling the phosphate spheres to the size $d_p=4.2$\AA.
A grooved structure, which resembles the real DNA appearance, is achieved in
the MAM \cite{montoro1998} through the adding another neutral sphere between the
cylindrical core and the charged 
phosphate sphere. The cylindrical core in the MAM has a diameter
$D=7.8$\AA, \, the inner string of neutral
spheres is centered at a radial distance $r=5.9$\AA, \, 
 and the outer string
of phosphates is centered at a radial distance $r= 8.9$\AA. 
 Both spheres have the same $\phi$ and $z$ coordinates and diameter $d_p=4.2$\AA.
A full description of these models is given in
Refs.~\cite{ourfirstDNApaper,oursecondDNApaper,montoro1998}. 
 
In addition to the two DNA molecules the system contains counterions 
of charge $q_c$, symmetric salt ions of
concentration $C_s$ and charges 
$q_+$ and $q_-$. 
All the small ions are modelled as a hard spheres of a diameter $d_c$ for
counterions, $d_+$ and $d_-$ for the salt ions. 
The whole system is held at room temperature $T=298K$.
 The primitive model  simulations with no explicit water deal
 only with a passive (non-specific) binding and completely 
neglect the specific  binding of counterions to the DNA grooves. In this case
 the ion binding sites
are determined by the steric and Coulombic interactions
 \cite{oursecondDNApaper,bonvin2000ebj}.

The interactions between the mobile ions and the phosphate charges 
are described within the framework
of primitive model as a combination of the excluded volume and Coulomb
interactions reduced by the inverse of the dielectric constant $\epsilon$ of the solvent.
The corresponding pair interaction potential between the different 
charged hard spheres is
\begin{equation}
V_{ij} (R) =\cases {\infty &for $ r \leq (d_i+d_j)/2 $\cr
   {{q_i q_j e^2} \over {\epsilon R}} &for $ R > (d_i+d_j)/2 $\cr}.
\label{1cLMH}
\end{equation}
where $R$ is an interparticle separation distance, $i$ and $j$ are indices denoting
the different particles species. Possible values for $i$ and $j$ are
$c$ (for counterions), $+,-$  (for positively and negatively charged
salt ions), $p$ (for phosphate groups) and $n$ (for the neutral
spheres in the MAM with $q_n$=0).
In addition, there is an interaction potential $V^{0}_{i}$ between the
DNA hard cylinder and the free ions
$i= c,+,-$. This potential has a simple excluded volume form such that
the free ions cannot penetrate into the cylinder.

\subsection{Simulated Quantities.}
Our basic simulated quantity is the effective
 force  \cite{ourfirstDNApaper,allah} between the DNA molecules
 \begin{equation}
{\vec F} = {\vec F_1} + {\vec F_2} + {\vec F_3}. 
\label{neu4}
\end{equation} 
Here ${\vec F_1}$ is 
the direct Coulomb force acting onto all the phosphate charges
belonging to one helical turn  of one DNA molecule as exerted from the 
phosphate groups of the other DNA,
\begin{equation} 
{\vec F}_1= -{\sum_k}^{'} \left( {\vec \nabla}_{{\vec r}_k^p}
  \sum_{n=1; n \not= k}^{N_p} V_{pp} \left( \mid {\vec r}_k^p - {\vec
      r}_n^p \mid \right)
\right).
\label{f1}
\end{equation} 
The sum $\sum_k'$ only runs over phosphates of one helical turn
of the DNA molecule. 

The second term ${\vec F}_2$ corresponds to the Coulomb
interactions between the phosphate charges and the mobile salt ions. 
This term describes the screening of the  DNA charge,
\begin{equation} 
{\vec F}_2= -{\sum_k}^{'} \left( \langle \sum_{i=c,+,-} \sum_{l=1}^{N_i}
 {\vec \nabla}_{{\vec r}_k^p}  V_{pi}( \mid {\vec r}_k^p - {\vec
 r}_l^i \mid ) \rangle \right). 
\label{f2}
\end{equation} 

The third term ${\vec F}_3$ arises from the entropic contribution of
small ions due to their excluded volume interaction with the DNA molecular
surface ${\cal S}_i$. Its value for one helical turn is
\begin{equation}  
{\vec F}_3=-k_BT \int_{{\cal S}_i} d{\vec f} \ \
\left(\sum_{j=c,+,-} \rho_j({\vec r}) \right ) ,
\label{9}
\end{equation} 
where ${\vec f}$ is a surface normal vector pointing outwards the DNA
cylindrical core. This term becomes increasingly
important as the Coulomb coupling parameter $\Gamma_{pc}$ is elevated
for the multivalent counterions  
\cite{ourfirstDNApaper,allah,Hartmut,wu1999}, 
\begin{equation}
\Gamma_{pc} ={\mid {q_p \over q_c }\mid}{2 \lambda_B \over
  {d_p+d_c}}.
\label{gamma_mc}
\end{equation}
The parameter $\Gamma_{pc}$ determines the importance of thermal fluctuations. When
 $\Gamma_{pc} > 1$, the Coulomb interaction energy between the DNA and
 the surrounding salt ions 
 dominates over the thermal fluctuations in system.

\section{Computer Simulations}
\label{simulations}
\begin{figure}
   \epsfxsize=13cm 
   \epsfysize=10cm 
~\hfill \epsfbox{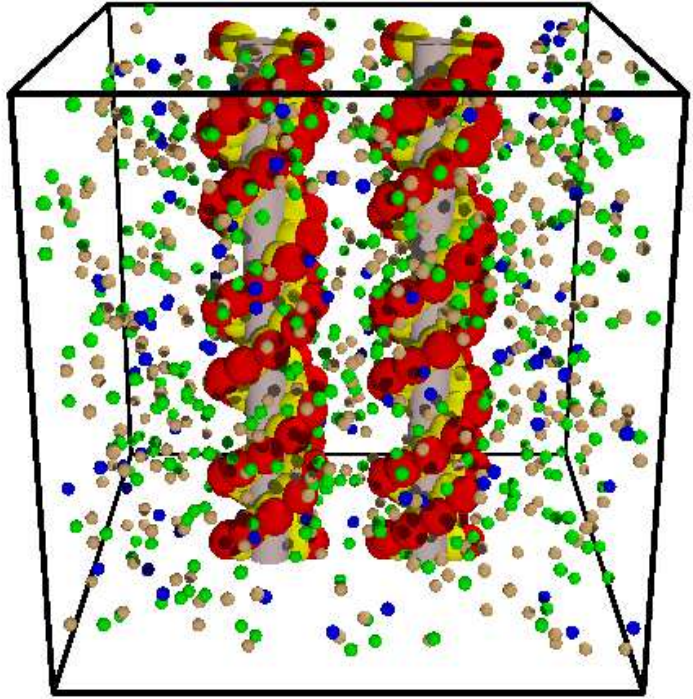} \hfill~
\caption{A typical snapshot of the simulation box. The DNA molecules are drawn
    according to the MAM. Black spheres on the DNA strands represent
    the phosphate charges. Internal grey spheres between the phosphates and
    the DNA cylindrical core are neutral. Positive (negative)
    salt ions spreaded across the simulation volume are shown as open 
(hatched) spheres.}   
\label{snapshot}
\end{figure}

We consider two  parallel DNA molecules, separated by a distance $R$
along the {\it xy} diagonal of the cubic simulation
box of size $L$ and volume $V=L^3$. The size of the simulation box
$L=102$\AA \, corresponds to the three 
full turns of B-DNA \cite{montoro1998}.
The box also consists $N_c$ counterions and
$N_+=N_-=N_s$ salt ions of both signs. 
The counterion concentration is fixed by the charge of 
DNA molecules in the simulation box due to the constraint
of global charge neutrality.
A typical snapshot of the simulation cell is illustrated in Fig.~\ref{snapshot}.
Periodic boundary conditions in all three
directions are applied to reduce the confined volume effects in
electrolytes. The DNA replicas in {\it z} direction produce an infinitely long
DNA molecule which avoids the end effects encountered in other
molecular simulations of short DNA segments
\cite{olmsted1989and1995,feig1999,allison1994}.  
 The phosphate spheres are monovalent, $q_p=-e$, where $e$ is an
 elementary charge. 
The total
number of simulated salt ions is varied from 0 to 2000 depending on the salt
concentration in the bulk.
Hereafter we will refer to each simulation by its nominal
salt concentration $C_s$ defined as a ratio between the total number of ions
$N_s$ in the cell without the DNA molecules and the system volume $V$, $C_s=N_s/V$. 
The dielectric permeability $\epsilon$ is considered to be a constant everywhere in
system, which avoids the need of electrostatic images \cite{hingerty,Tandon}.
 The long range interactions
between the two charged species and their 
replicas in the neighboring cells are handled via the Lekner method 
\cite{lekner} and its modification for the particular cases, when pair
of interacting charges are sitting exactly on one of the coordinate
axies (this case was considered in Appendix of Ref.~\cite{ourfirstDNApaper}). 

We have performed extensive molecular dynamics (MD) simulations for a
 range of different microion valencies. The simulated states are given in Table I.
The ion diameter was chosen to be $d_c$=3\AA. This
parameter defines the closest approach of the ion to the DNA surface
and has a strong impact on the polyion electrostatics
\cite{lyubartsev1995,stigter1996,fujimoto1994bj}. 
A test simulation for an increased ion diameter $d_c$=5\AA, which mimics the
ion hydration in solvent
 \cite{lyubartsev1995,oursecondDNApaper,conrad1988biopol,fujimoto1994bj},
 shows no qualitative changes of the reported 
results. We want to emphasize again that specific ion
 effects, as exemplified by the 
Hofmeister effect \cite{ninham1997langmuir,protein2}, 
are not accounted for in our model.

During the simulations, we calculate the interaction forces between the
two DNA molecules for 
different separation distances $R$. Due to the strong screening of the
DNA phosphates, the actual salt concentration in the bulk of the
simulation box $ C_s^{'}(R)$, measured far away from the polyelectrolytes,
 is $R$ dependent and is always smaller than the
nominal salt concentration $C_s$, $ C_s^{'}(R) < C_s$.   
 Thus, an implementation of the conventional MD procedure
 with a fixed ion number $N_s$
will yield to the interaction forces which correspond
to bulk densities $ C_s^{'}(R)$ for each
intermolecular distance $R$. This problem can be avoided by
considering a solution with a constant a chemical potential $\mu$ 
via the Grand Canonical (GC) simulation method \cite{kjellander1992chempot}.
The GC simulation is a natural choice to mimic the experimental
situation where the actual salt concentration of the ordered DNA phase is not
known {\it a priori}. Instead, it is given by the
thermodynamic condition that the 
chemical potential $\mu$ in the DNA solution has to be the same as in the bulk
electrolyte phase with which it is in equilibrium and whose
concentration $C_s$ is experimentally known.
 Thus the number of ions in the simulation cell is automatically
adjusted to the specified value of chemical potential $\mu$, which, in turn,
is linked to the concentration of ions in the bulk phase $C_s$
\cite{pettitt,lo1995,allen,lupkowski1991}.  
In the present paper a combination of different grand canonical
molecular dynamics (GCMD) schemes is used
which is optimally suited for our task.

\section{Grand Canonical Molecular Dynamics}
In addition to the usual propagation of the particles, the
conventional GC simulation technique \cite{allen,lupkowski1991}
consists of the
creation of particle at a random position in the simulation box or
destruction of a randomly chosen particle. Each of these moves is associated
with a probability of acceptance, which is determined by the ratio of two Boltzmann
factors. In application to electrolytes a modified GC method was
devised in
Refs.~\cite{vlachy,korolev,valleau1980,korolev1999,yau1994,torrie1980},
where the insertion or removal of a pair of ions of the same valency
and opposite charge is  
 done simultaneously to keep the system electroneutral. 
Unfortunately, these moves have relatively low acceptance
rates for a dense and multivalent salt solutions \cite{vlachy},
 making the simulations inefficient. 
Special methods, similar to the cavity-biased method
\cite{yau1994} and the gradual particle insertion method \cite{mon1985}
developed for uncharged systems, have
to be applied to overcome these obstacles in electrolytes. A biased
insertion/destruction procedure, apt for an application to low
temperature ionic fluids, is reported in Ref.~\cite{orkoulas1994}. 

Another challenge in the GC simulations is
the apparent incompatibility of the deterministic and
stochastic approaches. The dynamical information will be
adversely affected when particles suddenly appear and disappear.
This effect becomes even more pronounced for a non-homogeneous system
\cite{roij2000,roij1997and1999}, like
 a DNA immersed in solution, where an artificial and unrealistic
ion flow toward the DNA surface appears. This will further
destabilize the system equilibrium.
To minimize this inconsistency of the system dynamics, a method of local potential
control (LPC), first introduced in Ref.~\cite{papadopoulou1993}, can be
adopted. Within the LPC method, the creation and
destruction of particles is restricted to a control volume.
The other possibility is a procedure developed by Attard
in Ref.~\cite{attard1997}, where the GCMD is performed with a fixed number of
particles by coupling the variations in the system size to the
instantaneous chemical potential determined by the virtual test particle
method. This method also cures the low acceptance rates of
particle insertions and deletions for dense systems. 
 In the present simulations we take advantage of both the above mentioned methods
\cite{papadopoulou1993,attard1997}.
In detail, we first determine the specified nominal chemical potential
$\mu$ of the
bulk electrolyte in the absence of DNA molecules via a modified Widom method
with multiparticle insertion 
\cite{Svennson_gcmc,jonsson_lecture}. Then we match the actual chemical potential
$\mu^{'}$ to the nominal
chemical potential $\mu$ using a GCMD simulation similar to the method
invoked in
Ref.~\cite{papadopoulou1993} and locate the control volume near the cell
boundary. 
At each time step an equal but arbitrary number of creation/deletion
attempts are made in the control volume. After a successful creation,
a velocity is drawn from the Maxwell-Boltzmann distribution at a temperature
$T$ and assigned to the new particle. 
In the first stage the particle number $N_s^{'}$ in the simulation box
increases monotonically from its initial value
$N_s$ given in Table I. Then $N_s^{'}$ approaches its saturated value
and starts to fluctuate around it. This is followed by the
fluctuation of instantaneous chemical potential $\mu^{'}$ around the $\mu$.
At this stage we fix the particle number and allow the system size to fluctuate
according to procedure given by Attard in Ref.~\cite{attard1997}. The
fluctuations along the $x$ and $y$ 
directions ($z$ direction is strictly bound to the DNA length) never exceed a few percents of the box size $L$.
Our test simulations with and
without the Attard method \cite{attard1997} show the
equivalence of the algorithms, with  the former being much faster.

\section{mutual DNA configurations}
\begin{figure}
   \epsfxsize=9cm 
   \epsfysize=9cm 
~\hfill\epsfbox{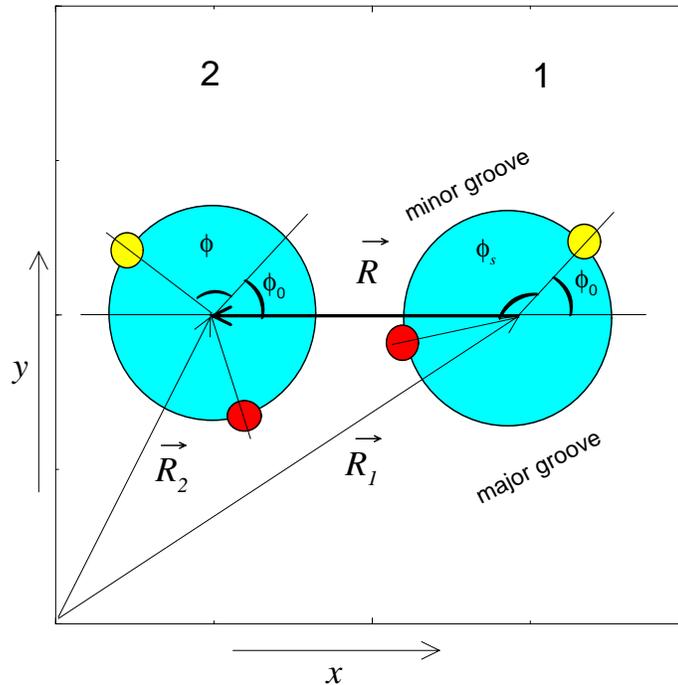}\hfill~
  \caption{A schematic picture explaining the positions of
     DNA molecules and the definition of the different azimuthal
     angles $\phi_s, \phi_0, \phi$. For further information, see text.}
 \label{xyplane}
\end{figure}

\begin{figure}
\hspace{-3cm}
   \epsfxsize=8cm 
   \epsfysize=4cm 
~\hfill\epsfbox{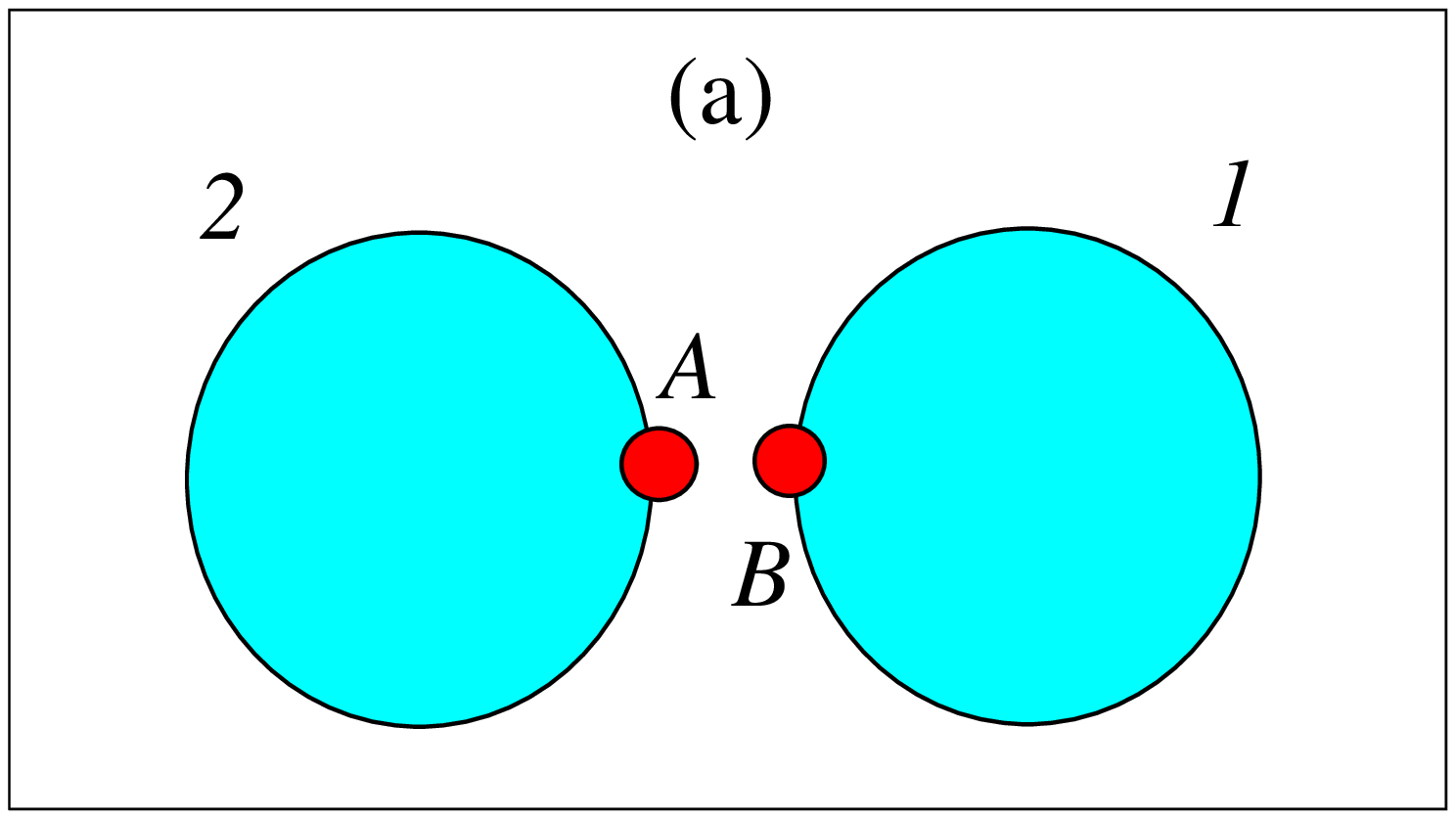}
   \epsfxsize=8cm 
   \epsfysize=4cm 
~\hfill\epsfbox{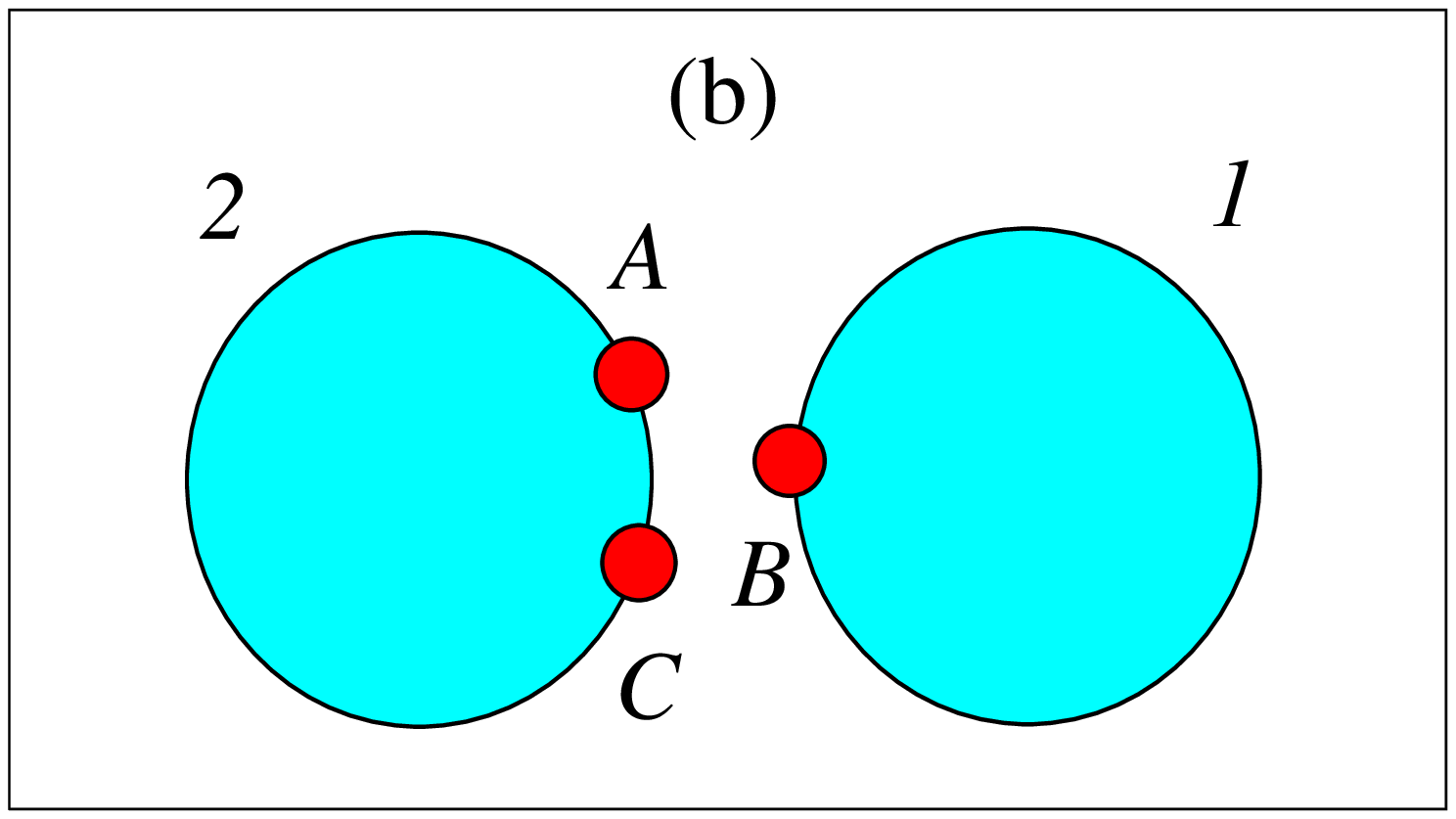}\hfill~
   \epsfxsize=8cm 
   \epsfysize=4cm 
~\hfill\epsfbox{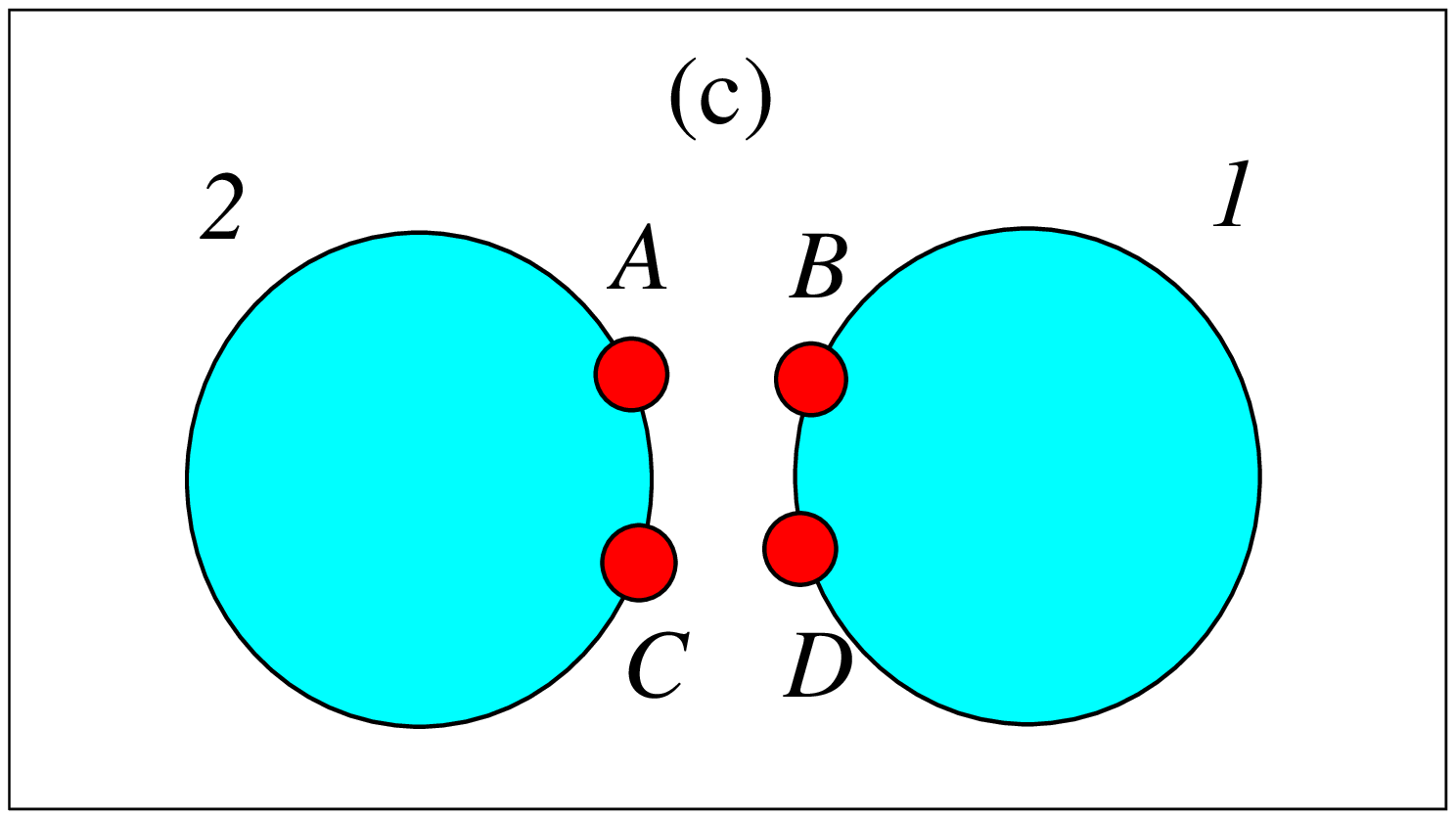}\hfill~
  \caption{Three typical configurations for the two parallel DNA molecules
    when their phosphates charges are close to each other. The {\it
     xy}-plane cross-sections of DNA molecules are shown for the ECM. Note
     that only the neighboring phosphates in the inter-DNA area are
     shown, see the dark
     small circles labeled by letters A, B, C, D. (a)
     $\phi_0+n\phi_s=\pi$, $\phi=\pi, \pi/5 $. (b) 
     $\phi_0+n\phi_s=\pi$, $\phi=\pi - \phi_s/2, \pi/5 - \phi_s/2$. 
(c) $\phi_0+n\phi_s=\pi \pm \phi_s/2$, $\phi=\pi, \pi/5$. Here $n$
     is an integer number, $n=0, \pm 1, \pm 2, ...$. In (b) and (c)
     the pair of phosphates on each
     cylinder pertain to the same strand and have different $z$
     coordinates: (b) $z_A=z_B-1.7$\AA, $z_c=z_B+1.7$\AA; (c)
     $z_A=z_D$, $z_B=z_C$, $z_A-z_c=3.4$\AA.   
}
 \label{configure}
\end{figure}

We calculate the total interaction force $F (R)$, Eq.(2), and its components
 $F_2(R)$ and $F_3(R)$, compare Eqs. (4) and (5), respectively,
 for a given nominal salt concentrations
 $C_s$. The direct phosphate-phosphate interaction $F_1(R)$, Eq.(3), does not
 depend on the salt density and its assessment is straightforward. 
It should be mentioned that in addition to the separation distance
 $R$, there are three  angular variables which define the mutual
 configuration of two DNA
 molecules. These variables,
 the azimuthal angles $\phi_s$, $\phi_o$ and $\phi$ are shown in
 Fig.~\ref{xyplane}. 
The angle $\phi_s$ defines the widths of the DNA
 grooves \cite{oursecondDNApaper} in the {\it xy} plane, it is $134^o$ for the CM and ECM,
 and $154^o$ for the MAM. The parameter $\phi_0$ is the angle between the 
phosphate charge and the DNA-DNA separation vector $\vec R=\vec R_1 - \vec
 R_2$ and characterizes the discrete 
location of the phosphate charges along the strands. All
 the results for the DNA-DNA interaction are periodic in $\phi_0$ with a periodicity of
 $36^\circ$. The angle $\phi$ describes the rotation of the
 second DNA cylinder around its long axis with regard to the first
 DNA cylinder.
There are five particular DNA-DNA configurations which make a strong
 contribution to the interaction force at short separation
 distances \cite{ourfirstDNApaper}. Three of these configurations
 correspond to the case when the
 phosphate charges of neighboring DNA molecules are ``touching'', see
 Fig.~\ref{configure}. The other two particular
 configurations correspond to the so called ``DNA
 zipper''situation, when the strands of one DNA stand against the
 grooves of the neighboring DNA \cite{zipper}. This happens when  
 $\phi=\pm 3\pi/5$ regardless the value of $\phi_0$. 
Our previous \cite{ourfirstDNApaper} and present simulations
prove that the interaction force  does depend on the mutual DNA
configurations at a short separation distances $R<25$\AA,\, or when the
surface-to-surface distance between the two DNA molecules is less than 5\AA. This
follows mainly from ion bridging between the two neighboring phosphates
via a positive salt ion in configurations pictured in
Fig.~\ref{configure}, or from sharing an adsorbed salt ion in one DNA groove
and on the other DNA strand in the DNA zipper configurations. A short range
 attractive interaction between the charged rods arising from 
 interlocking counterions, also sometimes called counterion
cross-links,  between the rods has been investigated in
 Ref.~\cite{deserno2002attraction,schellman1984crosslink,linse2002attraction}. 
In fact,  for  such small separations, comparable with the solvent
particle size, discrete 
solvent effects will show up {\it in vitro}
\cite{schneider1998,kjellander2001solvent}.  
On the other hand, the multivalent ions increase
the hydrodynamic radius of the DNA molecule, which in turn makes it unlikely
for two neighboring DNA molecules to come closer than the contact
shell-to-shell distance \cite{fujimoto1994bj}. 
Arguments against the existence of
cross-links for multivalent ions are given in Ref.~\cite{pelta1996}
in order to explain the fluidity of the condensed phase of DNA system.
X-ray scattering experiments in DNA aggregates a
show that the surface-to-surface 
spacing between the DNA helices is only about one or two water molecule diameters
\cite{schellman1984crosslink}.
Thus, numerical results for small DNA separations,
accessible in simulations but subject to a complicate
statistical averaging procedure, bear no physical meaning
 to match the experimental results. 
For larger
separation distance, $R>25$\AA \, we find no detectable dependence of the
interaction forces on the azimuthal angles $\phi_0$ and
$\phi$.  This is in accordance with the early reports
\cite{wagner,lyubar,montoro1998,gulbrand1989,jaya2,conrad1988biopol,hochberg}
that the  helicity and discreteness effects of the DNA charges are
generally small and dwindles a few angstroms away from the DNA
surface. In all figures hereafter
we show the interaction forces starting from the distances $R=24$\AA. 
The interaction forces are scaled per
DNA pitch, i.e. per 10 DNA base pairs. A positive
sign of the forces denotes a repulsion, while a negative sign denotes
an attraction. The cases of monovalent and multivalent salt ions are
considered separately. 

\section{Simulation results for monovalent salt.}
\subsection{Monovalent Counterions.} 
\begin{figure}
\hspace{-3cm}
   \epsfxsize=9cm 
   \epsfysize=9cm 
~\hfill\epsfbox{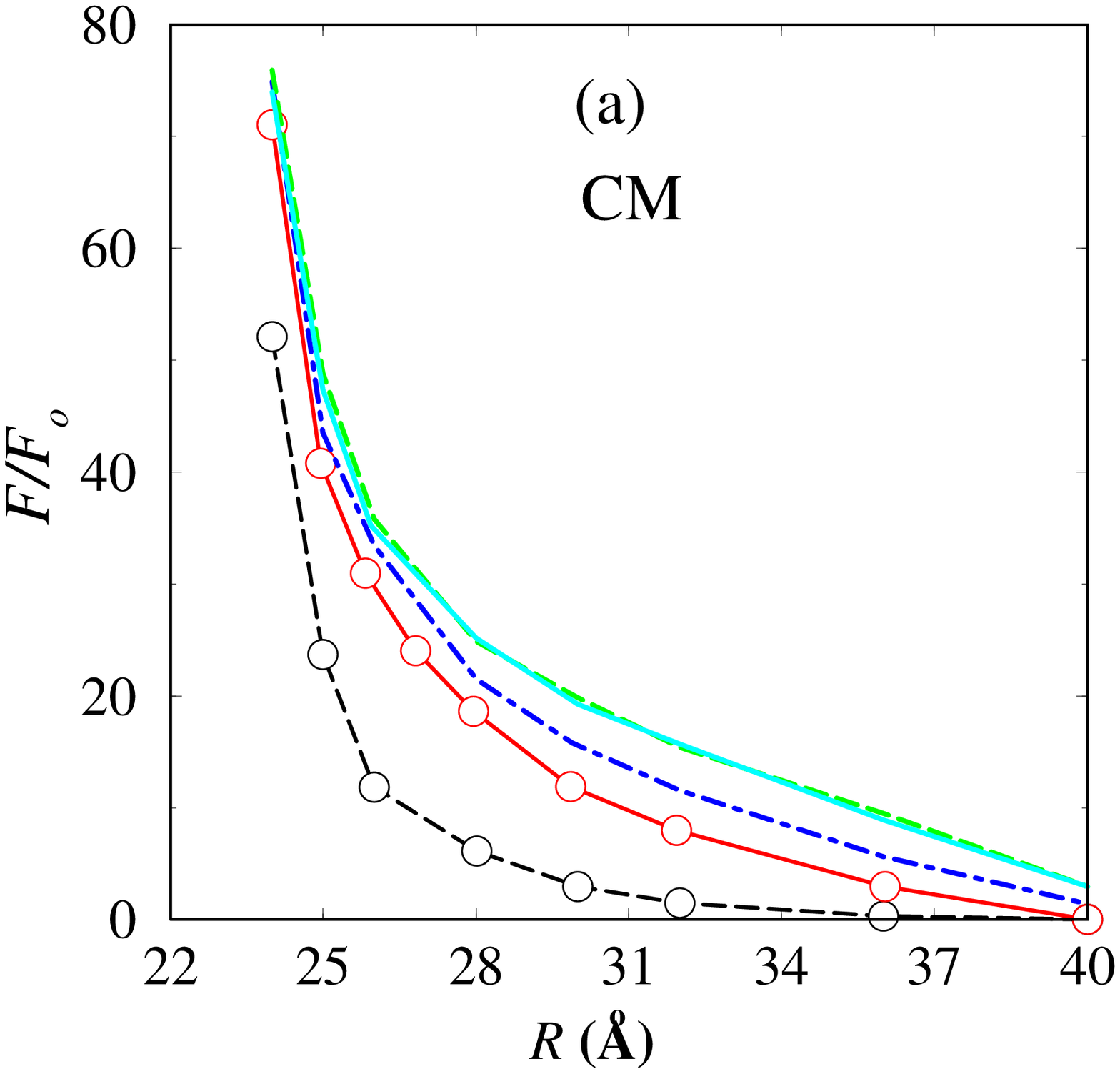} 
   \epsfxsize=9cm 
   \epsfysize=9cm 
~\hfill\epsfbox{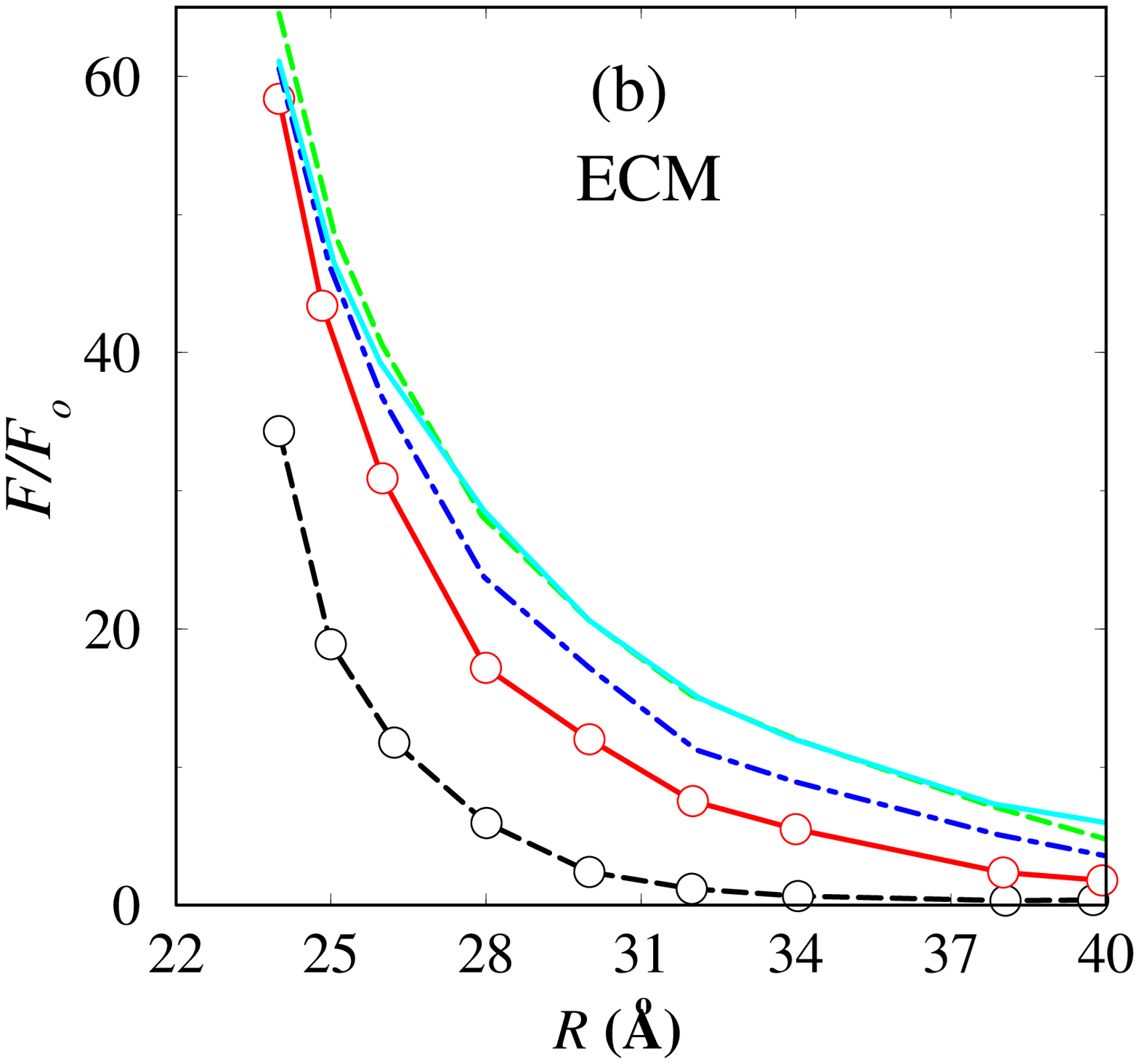}\hfill~
   \epsfxsize=9cm 
   \epsfysize=9cm 
~\hfill\epsfbox{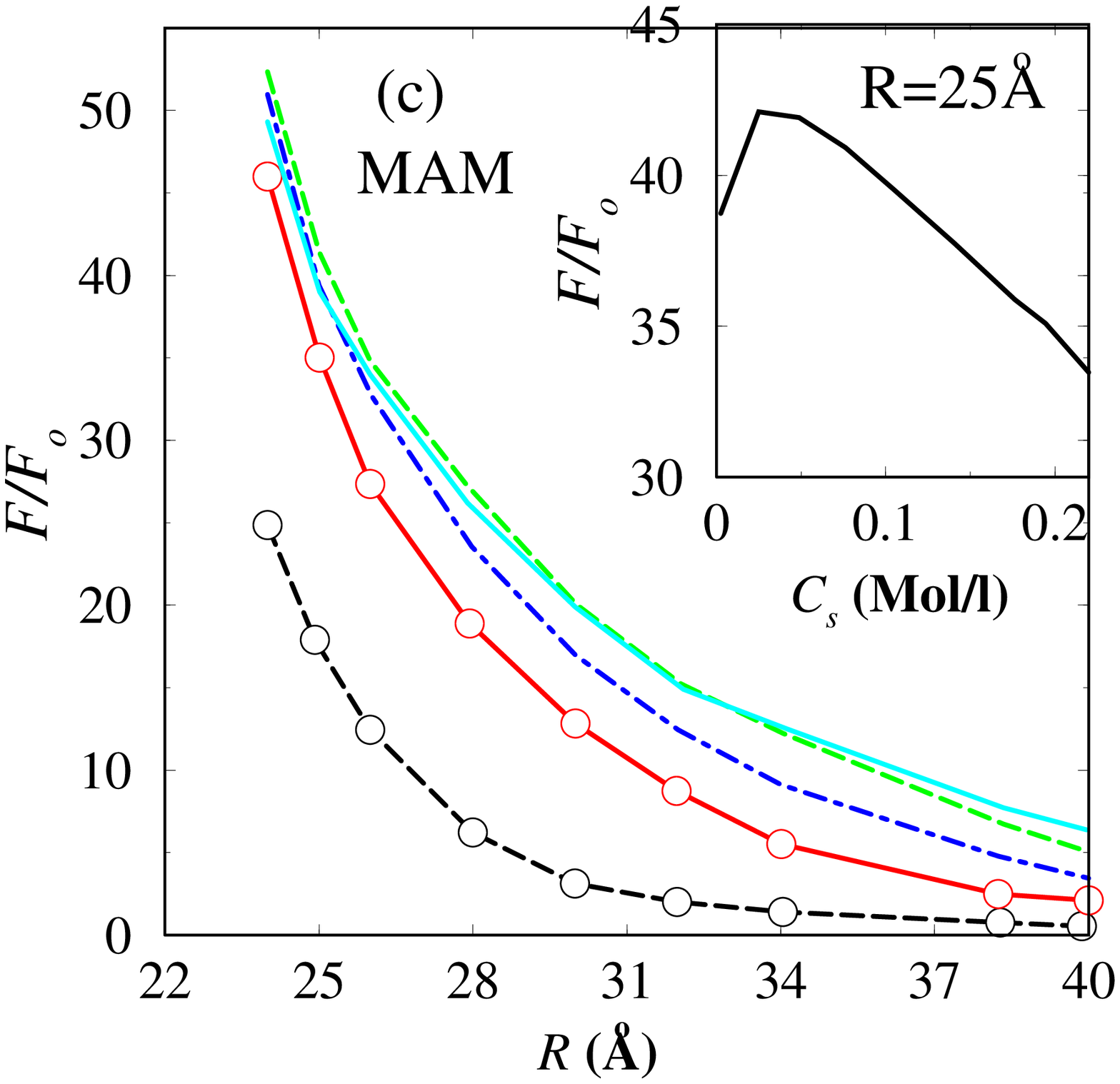} 
\caption{Reduced DNA-DNA interaction force $F/F_0$ versus
  separation distance $R$ for
  the monovalent counterions and monovalent salt ions 
(parameter set 1 of Table I).
  The unit of the force is $F_0=k_BT/P$, where $P$ is the DNA pitch length. 
  Different salt densities are shown: $C_s$=0 mol/l (solid line), 0.024 
  mol/l (dashed line),  0.097 mol/l (dot-dashed line),  0.194mol/l
  (solid line with symbols), 0.71
  mol/l (dashed line with symbols). (a)- CM, (b)- ECM, (c)- MAM. The inset in (c)
  shows the force-salt non-monotonicity at the separation distance $R$=25\AA.} 
 \label{force_total_1_1}
\end{figure}

\begin{figure}
\hspace{-3cm}
    \epsfxsize=9cm 
    \epsfysize=9cm 
~\hfill\epsfbox{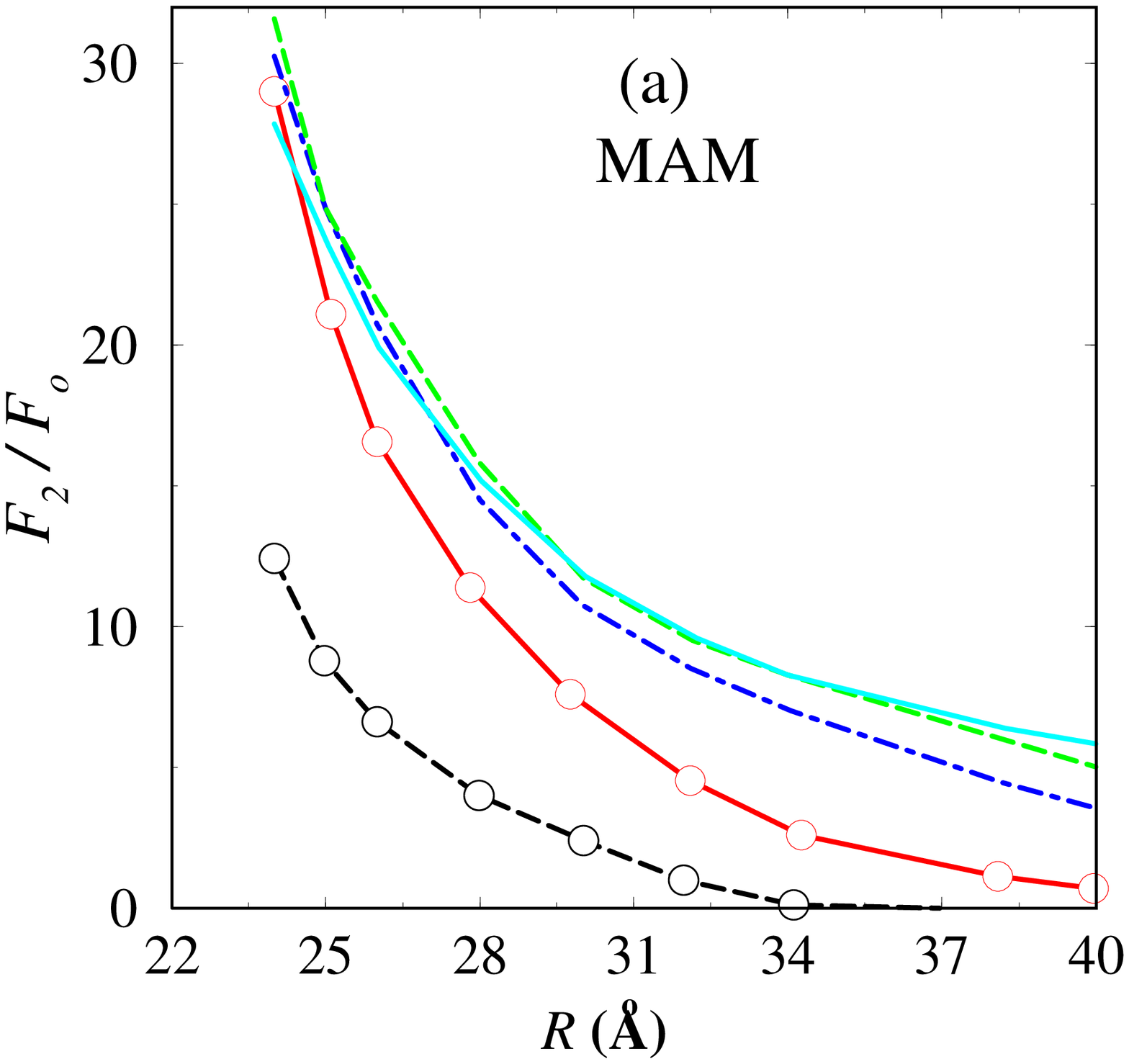}
    \epsfxsize=9cm 
    \epsfysize=9cm 
~\hfill\epsfbox{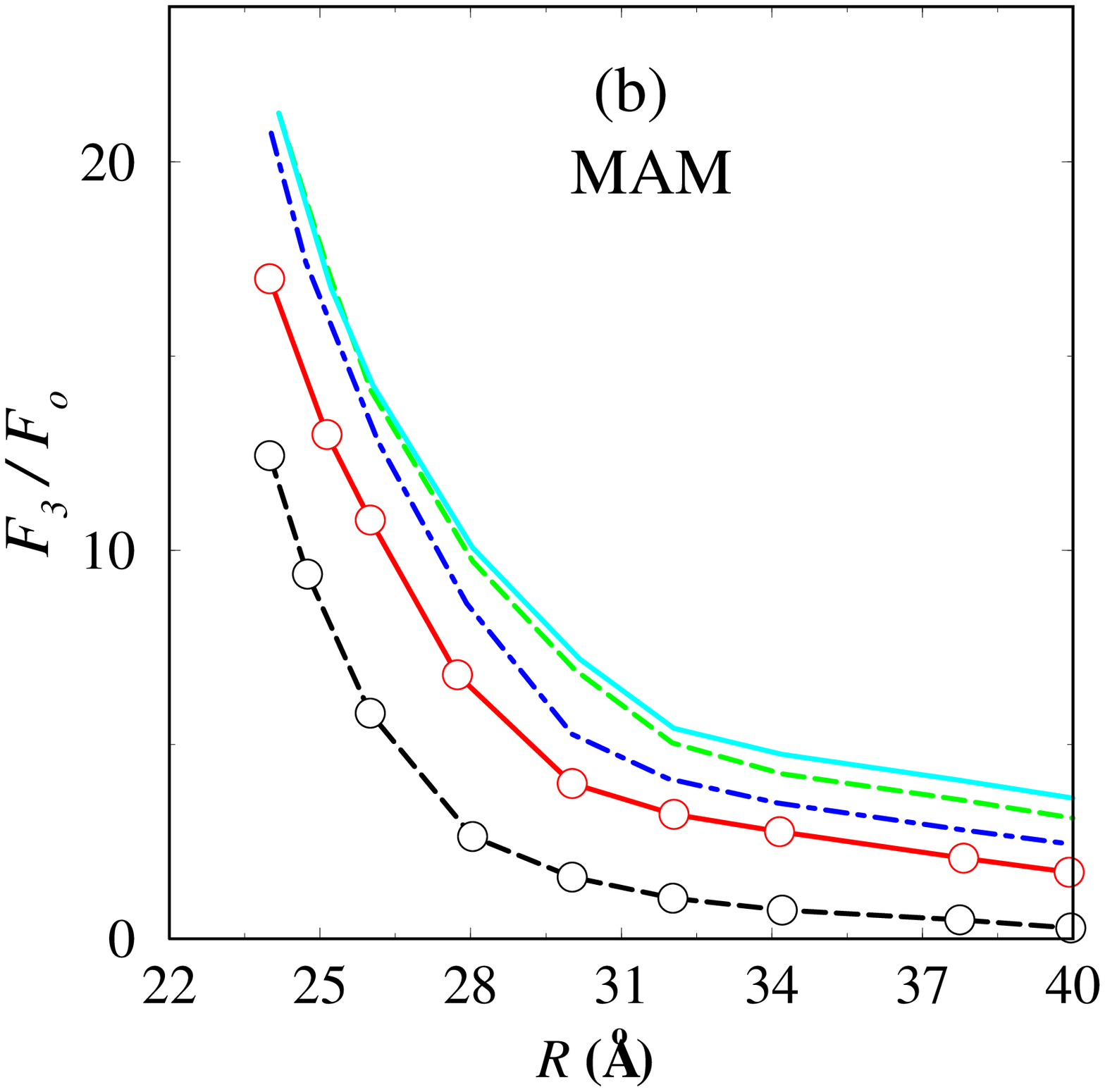}
\caption{(a) Reduced electrostatic  force $F_2/F_0$ and (b) entropic
  force $F_3/F_0$ components of the total
  interaction force $F(R)$ for the MAM, compare Fig.~\ref{force_total_1_1}c. 
A similar trend is observed for the CM and ECM.}
 \label{force_compo_MAM_1_1}
\end{figure}

The calculated DNA-DNA interaction forces for monovalent salt and
counterions are depicted in Fig.~\ref{force_total_1_1}. 
 All three DNA models exhibit a repulsion between the DNA molecules for all
 calculated separation distances
and salt densities shown in Fig.~\ref{force_total_1_1}
\cite{bloomfield1991biopolymers,ourfirstDNApaper}. The 
repulsion in the CM is
roughly twice as strong as in the MAM. This is
a result of grooved nature of the MAM where the vast majority of
adsorbed counterions sits in the minor groove \cite{oursecondDNApaper}.
The salt dependence of the force at a fixed, small separation appears to
be non-monotonic. This is clearly visible only for the ECM and MAM
but hardly detectable for the CM. In detail, if the
salt molarity is increased from  $C_s$=0
mol/l to $C_s$= 0.024 mol/l, the repulsion at short distances
becomes stronger, see the inset of
Fig.~\ref{force_total_1_1}c. Though this trend is at odds with the classical
screening theories, a similar effect has already been reported in
Ref.~\cite{ourfirstDNApaper,wu1999}.
A detailed consideration of the interaction force components in
Fig.~\ref{force_compo_MAM_1_1} reveals that
the non-monotonicity of the interaction force $F(R)$ has a purely electrostatic
origin. Both the electrostatic
 force $F_2(R)$ and the entropic
 force $F_3(R)$ are repulsive for
all indicated salt densities; the non-monotonicity is only
contained in $F_2$.
The non-monotonicity is a tiny effect for monovalent ions, but
for the divalent counterions and monovalent salt, this non-monotonicity
aggravates and induces a switch from attraction to repulsion, see the next subsection.

\subsection{Divalent Counterions.} 
\begin{figure}
\hspace{-3cm}
   \epsfxsize=9cm 
   \epsfysize=9cm 
  ~\hfill \epsfbox{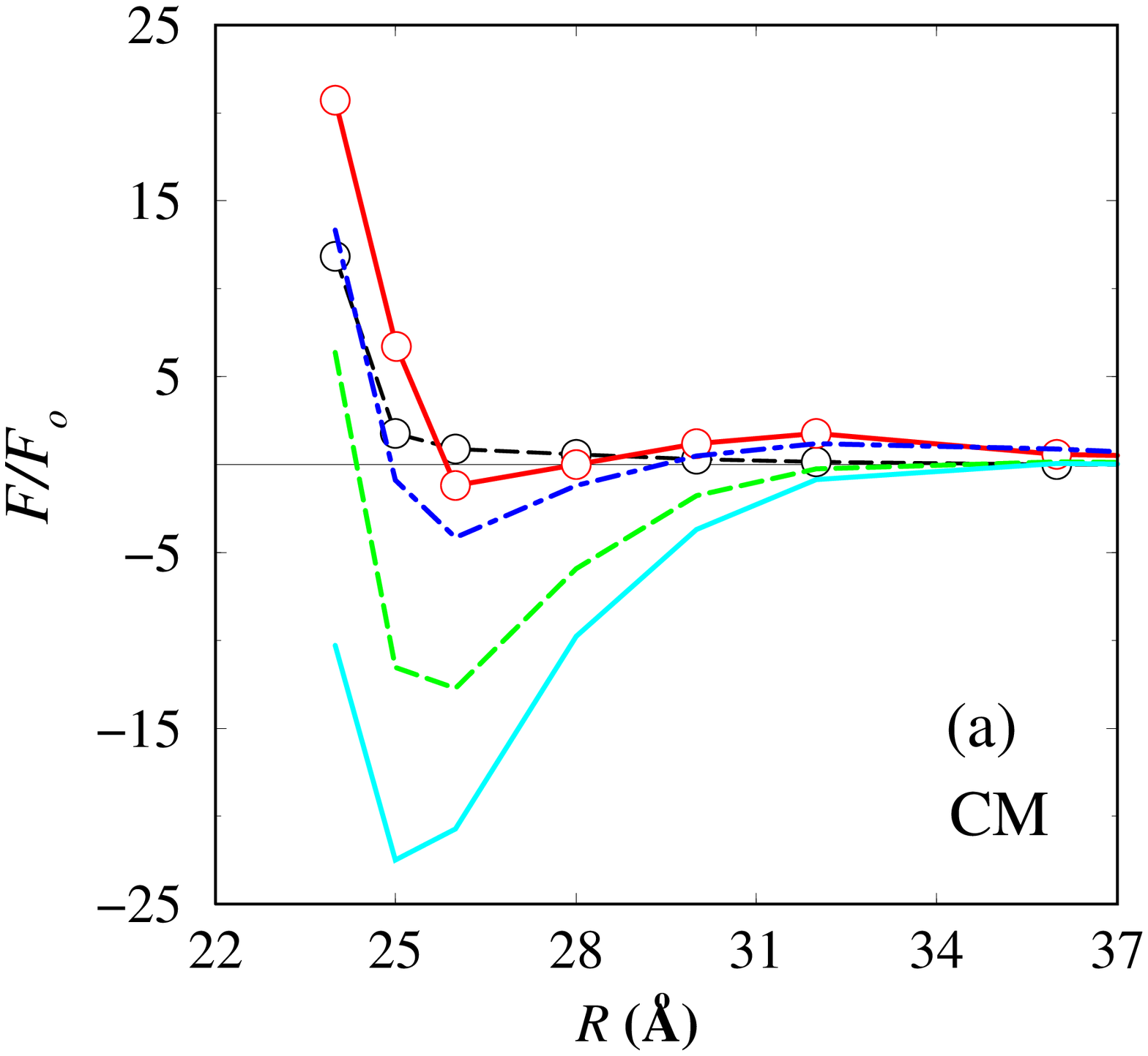} 
   \epsfxsize=9cm 
   \epsfysize=9cm 
  ~\hfill \epsfbox{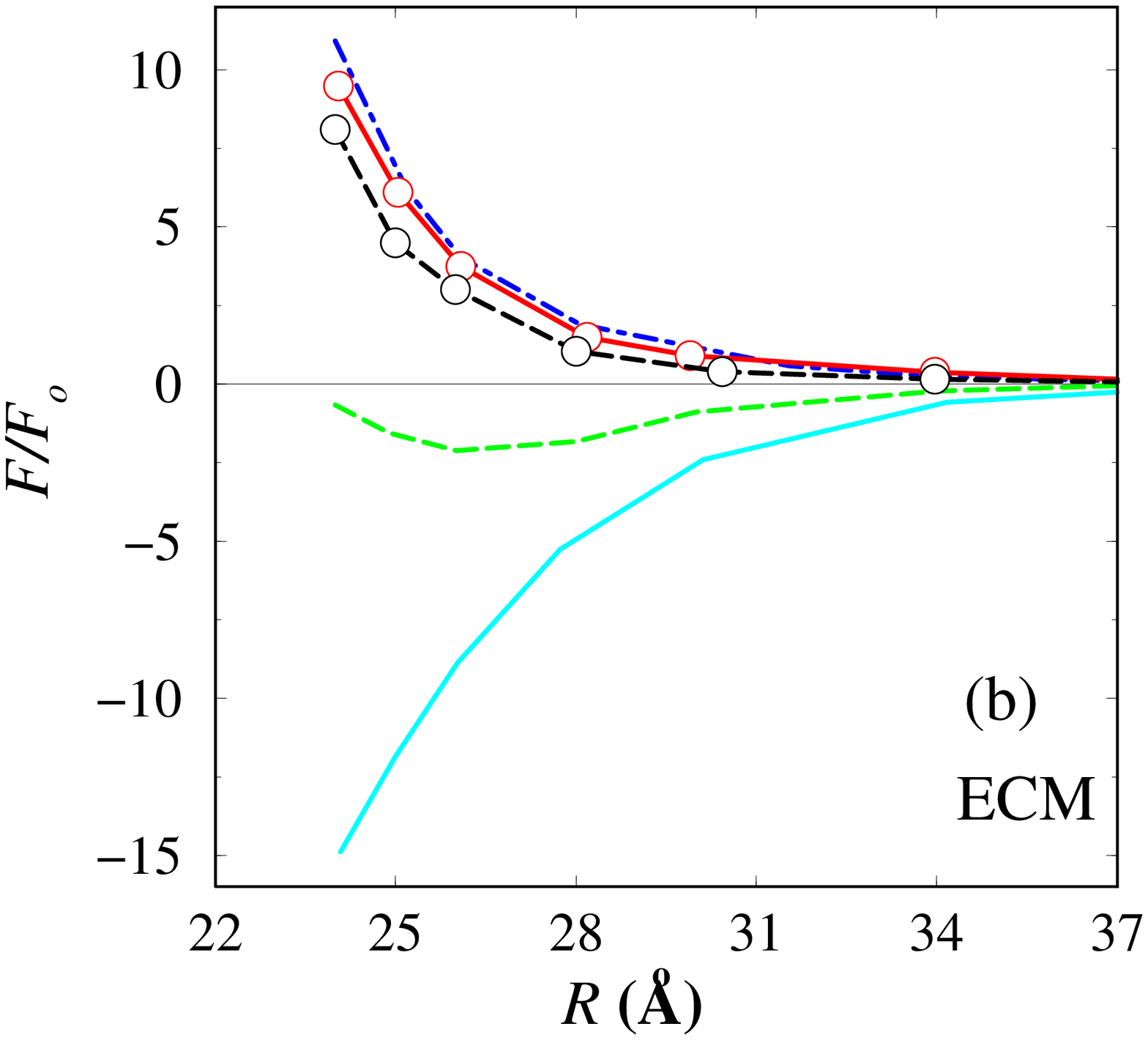} \hfill~
   \epsfxsize=9cm 
   \epsfysize=9cm 
 ~\hfill  \epsfbox{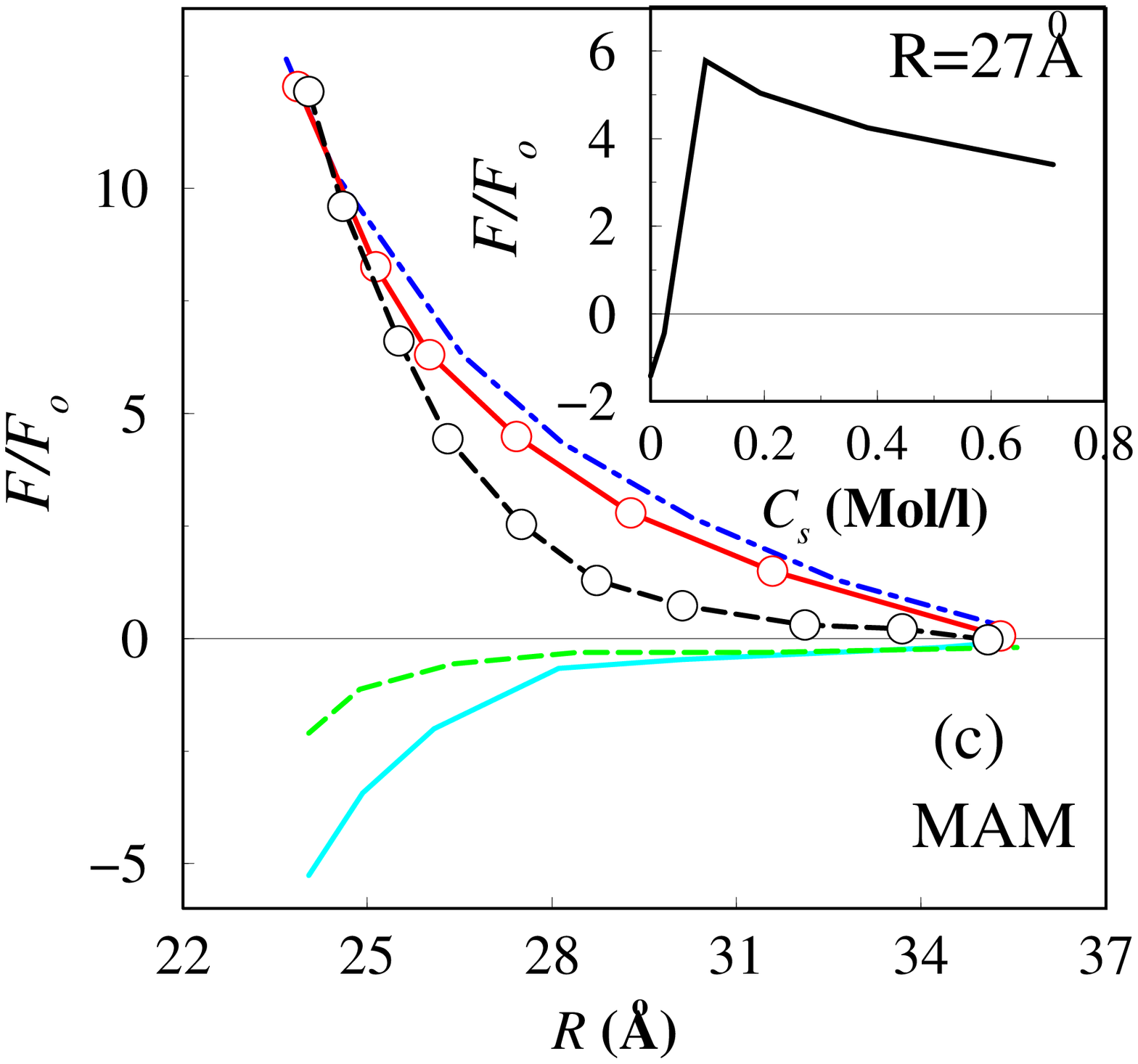}
\caption{Reduced DNA-DNA interaction force $F/F_0$ versus
  separation distance $R$ for
   divalent counterions and monovalent salt ions (parameter set 2 of Table I). The
  notation is the same as in Fig.~\ref{force_total_1_1}. The inset in (c)
  shows the force-salt non-monotonicity at the separation distance $R$=27\AA.}
 \label{force_total_2_1}
\end{figure}

\begin{figure}
\hspace{-3cm}
   \epsfxsize=9cm 
   \epsfysize=9cm 
~\hfill\epsfbox{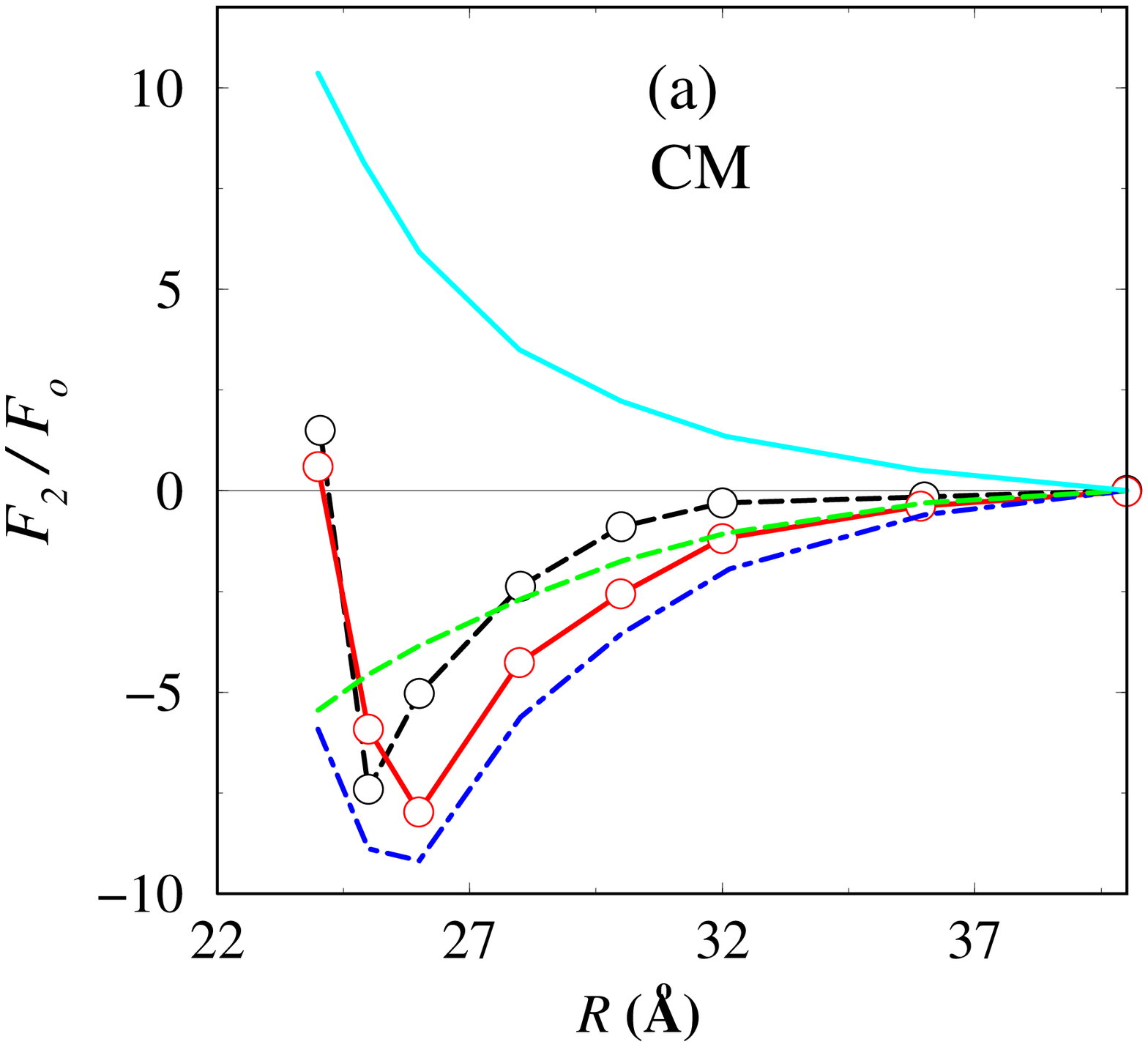}
   \epsfxsize=9cm 
   \epsfysize=9cm 
~\hfill\epsfbox{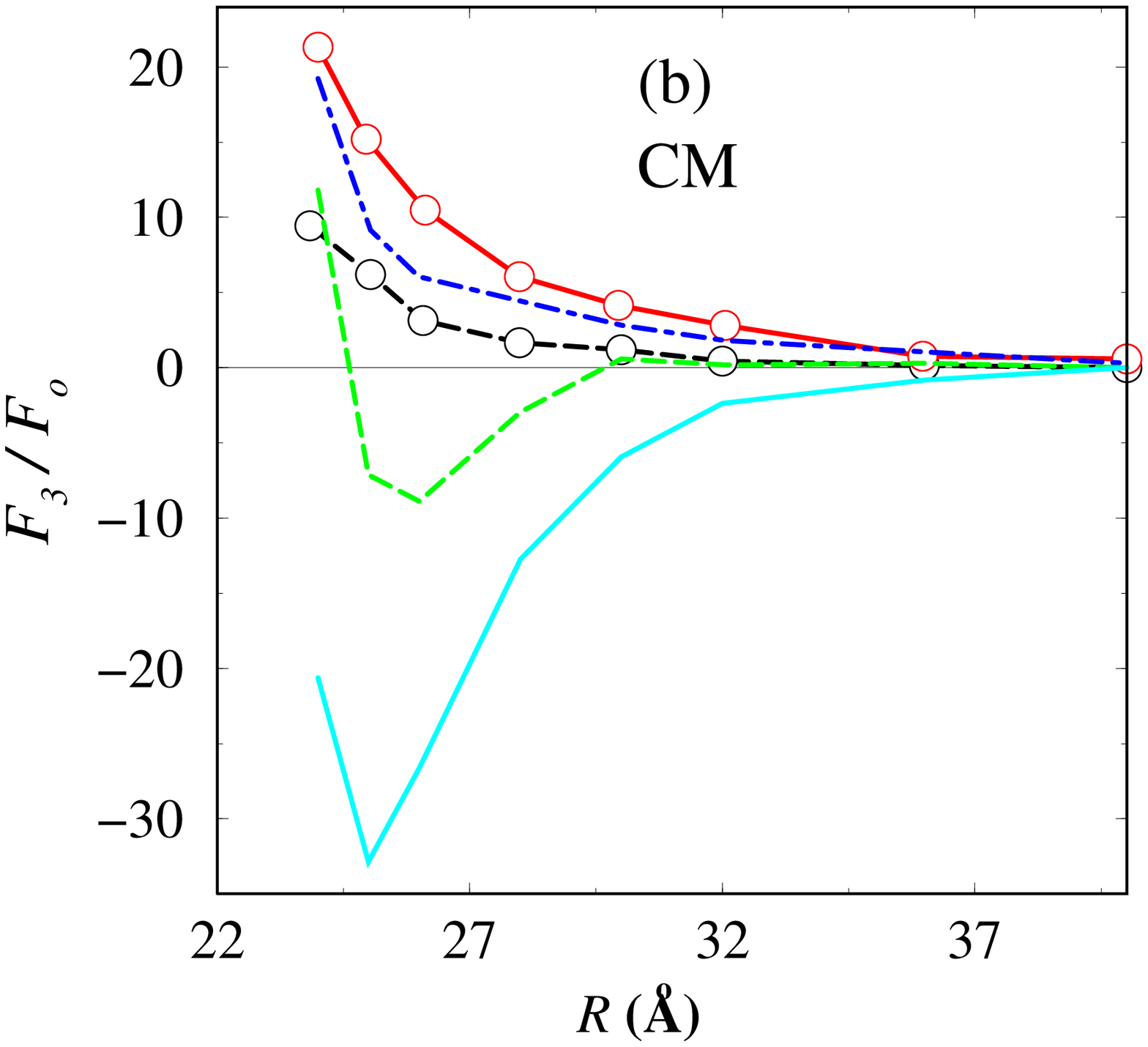}\hfill~
\caption{(a) Reduced electrostatic  force $F_2/F_0$ and (b) entropic
  force $F_3/F_0$ components of the total
  interaction force $F(R)$  from the Fig.~\ref{force_total_2_1}a for
  the CM. 
 The notation is the same as in Fig.~\ref{force_total_1_1}. }
\label{force_compo_CM_2_1}
\end{figure}

 The interaction forces for divalent counterions and
 monovalent salt, given in Fig.~\ref{force_total_2_1}, reveal
 a DNA-DNA attraction for small added salt concentrations.
First we analyze the salt free case, when the attraction in the CM is
 nearly three times stronger than in the MAM.  
The origin of these attractions in the CM and MAM is completely different. 
In the cylindrical model the attraction is totally associated with the 
"Coulomb depletion'' effect \cite{allah,HansenLoewen}. Such an ion
depletion is related to the
formation of strongly correlated counterion
 liquid  on the DNA surface. According to the results of
 Ref.~\cite{oursecondDNApaper}, the dense spots of counterion liquid 
are mostly associated with the DNA phosphates. 
For short DNA-DNA separations, when the mean separation
 distance between counterions on the DNA surface exceeds the
separation distance $R$, the two strongly correlated counterion
clouds on the different DNA rods start to repel each other. As a
 result, the 
local density of ions in the inter DNA area becomes less than its value
on the outside area. This in turn leads to a unbalanced
pressure from the outer counterions \cite{hydration_attraction}, which
 implies an attraction, as
 proven by the solid lines in Fig.~\ref{force_compo_CM_2_1}.
It is worth to mention that in
Ref.~\cite{allah} such a correlation mechanism was reported for a spherical colloid with
a central charge in a low dielectric medium. Surprisingly, for the cylindrical
macroions with a discrete surface charge considered here, we recapture a similar
effect. Note that our results do not support the
implications of Ref.~\cite{manning1992}, 
where an attractive force between the two rodlike polyions is assumed
 to be mediated by the
sharing of condensed counterions.

\begin{figure}
\hspace{-3cm}
   \epsfxsize=9cm 
   \epsfysize=9cm 
~\hfill\epsfbox{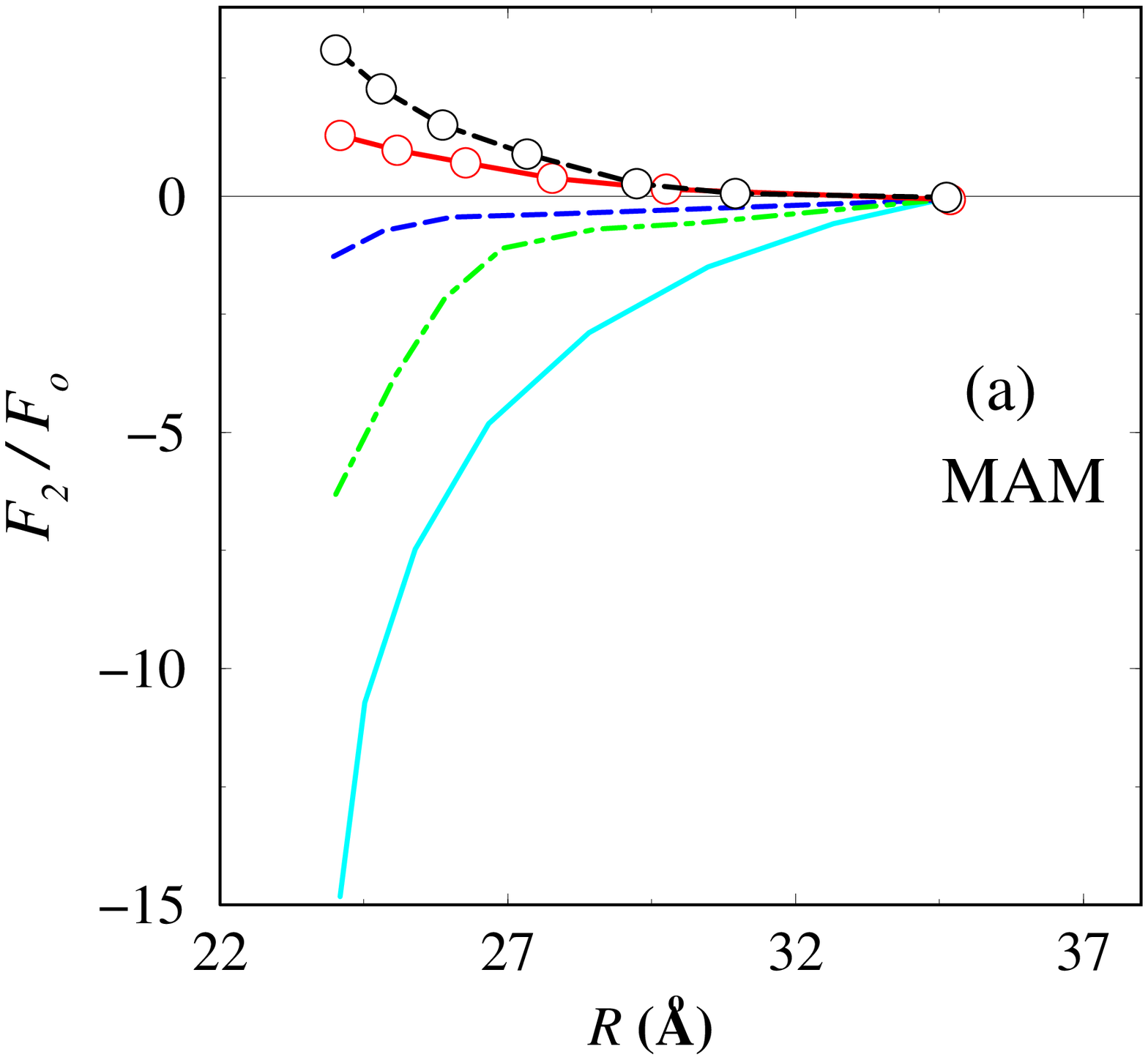}
   \epsfxsize=9cm 
   \epsfysize=9cm 
~\hfill\epsfbox{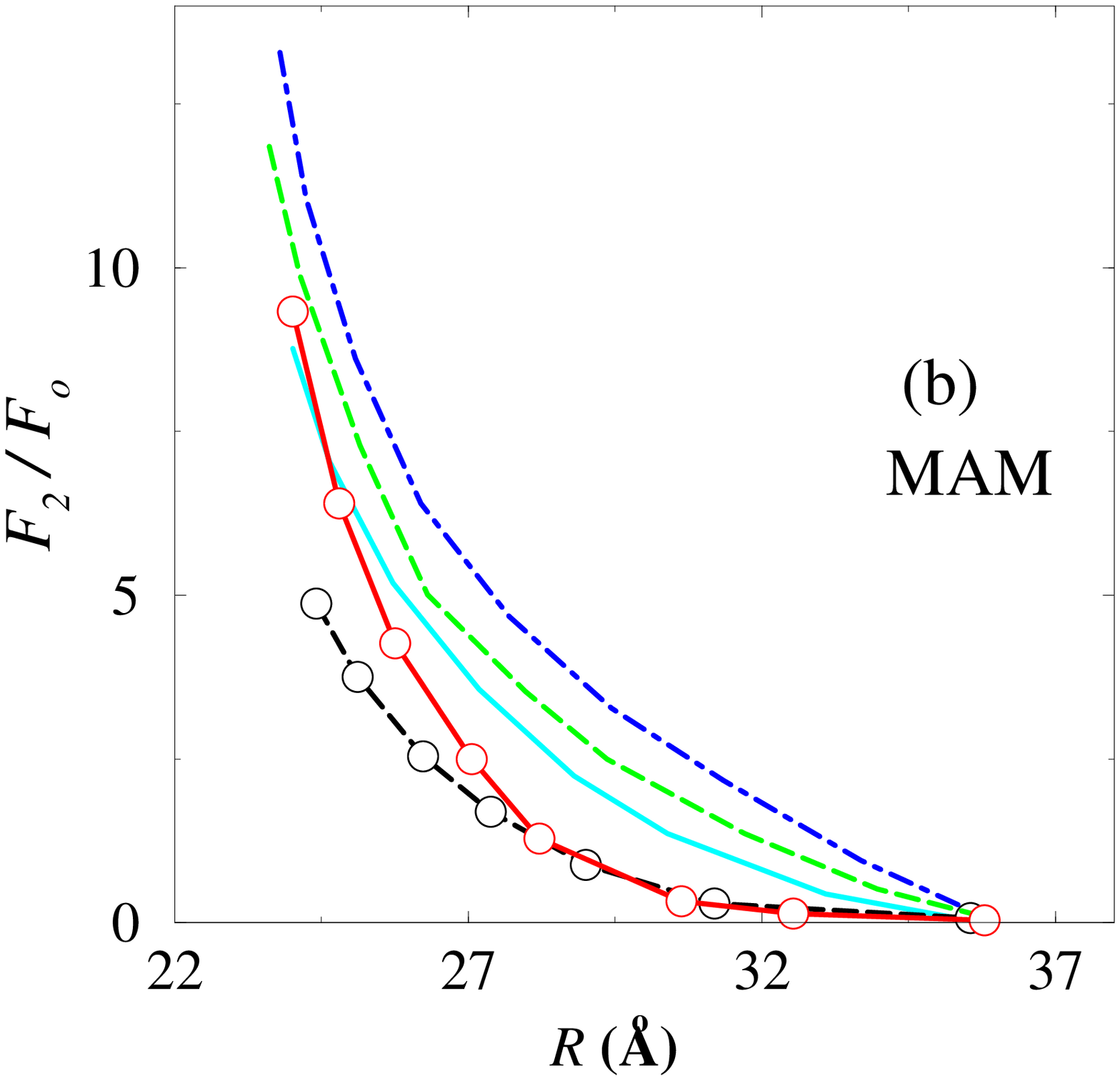}\hfill~
\caption{(a) Reduced electrostatic  force $F_2/F_0$ and (b) entropic
  force $F_3/F_0$ components of the total
  interaction force $F(R)$  from the Fig.~\ref{force_total_2_1}c for
  the MAM. The notation is the same as in Fig.~\ref{force_total_1_1}. }
\label{force_compo_MAM_2_1}   
\end{figure}

Contrary to the CM, the attraction in the MAM  and ECM has a purely electrostatic
origin, as proven by Fig.~\ref{force_compo_MAM_2_1}. There are more ions in
 the inter DNA-DNA area compared to the outer DNA-DNA area.
 The range and strength of this
 attraction is higher for the ECM than for the MAM. To our believe,
 this is due to the different counterion condensation patterns on
 the DNA surface in the MAM and ECM. In the latter model the
 ions predominantly adsorb to the DNA strands and in the minor
 groove. Therefore they occupy more DNA surface area compared to the MAM,
 where the main destination of the ion adsorption is the minor groove of
  DNA \cite{oursecondDNApaper}. According to a simple intuitive
  picture, the adsorbed divalent counterions 
form a strongly correlated fluid on each DNA surface. 
 A mutual arrangement of these two neighboring shells
with a minimal potential energy gives rise to the attractive
electrostatic force between the DNA molecules. 

Upon an addition of monovalent salt, the trend in the interaction
force depends sensitively on the DNA shape which is modelled
differently in the CM, ECM and MAM. 
 In the CM the DNA-DNA attraction persists to high
salt concentrations. Only at short separations and dense salt the
interaction has a repulsive branch. However, there is a counterintuitive
behavior of the force-salt dependence
 in the ECM and MAM: a small increase in the salt
 concentration $C_s$ suppresses 
 the attractive interaction force $F$. As $C_s$ increases further, the
 DNA-DNA interaction force becomes strongly repulsive
 over a broad range
 of separation distances. At even
 higher $C_s$, the interaction force $F$ is
 completely screened out and descends toward zero in
 accordance with the classical double layer theories. We denote this
 trend {\bf salt-induced stabilization}. It is in complete contrast to
 salt-induced destabilization or salt-induced coagulation which is
 typical for charged colloids \cite{PuseyLH,allahprotein,moon2000salteffect}.  
 The force-salt dependence at a fixed separation distance is shown in
 the inset of the Fig.~\ref{force_total_2_1}c. 
 To explain the mechanism of salt-induced stabilization, we separately
 plot the interaction force components in 
Fig.~\ref{force_compo_CM_2_1} and
Fig.~\ref{force_compo_MAM_2_1} for the CM and MAM, respectively. 
As the salt density is increased, the
entropic force in the CM first changes its sign from an attraction to 
a repulsion, then falls back to zero from above. The electrostatic
force in the CM behaves in a similar manner but with opposite sign. It first
goes down from the positive to the negative values, then approaches
zero from below. Hence, at higher salt densities the attraction,
seen in Fig.~\ref{force_compo_CM_2_1}, is electrostatically driven.

Contrary to the CM, the non-monotonicity of the interaction force in the ECM and
MAM has an entropic origin, see Fig.~\ref{force_total_2_1} and
Fig.~\ref{force_compo_MAM_2_1}. For a fixed separation
 distance $R$ it first goes up and then drops back to zero as
  $C_s$ is increased. Similar to the monovalent counterion case, the
  entropic force is repulsive over the full range of separation distances.  
 The electrostatic force, which  was totally repulsive for the monovalent
 counterions, now is attractive for dilute salt densities. Thus, it is the
electrostatics that makes the total interaction attractive for the ECM and MAM 
 at small salt densities.  
As $C_s$ is increased, the  monovalent ions tend to replace
the adsorbed divalent ions on the DNA surface in accordance with the two-variable theory of
Manning \cite{manning1978salt}. This replacement is
energetically favorable, since the divalent ions gain more polarization
energy in the bulk electrolyte. This results in the loss of the
attractive electrostatic force and the weakening of the entropic force. 
 At the final stage, when all the divalent sites on the DNA surface are occupied by
monovalent ions,  the strongly correlated
fluid structure is destroyed and the entropic and
electrostatic forces drop to zero.

An other interesting observation is the occurrence of zero force in the CM in
Fig.~\ref{force_total_2_1} at small separation distances. As the salt
concentration is increased, the distance where the interaction force
vanishes, shifts towards larger values. The physical meaning
of this effects is not clear to us. Probably it is an artifact of the
simple DNA model with no grooves. 

 We note that the salt densities invoked in the
current simulations are below the concentrations where a salt depletion
\cite{allahprotein} appears near the DNA surface. Thus the observed
attractive DNA-DNA force is not induced by the salt depletion effect,
which was claimed in Ref.~\cite{allahprotein} to be one of the main
contributors to the protein-protein attraction. Furthermore the attraction observed here
is unrelated to metastable ionization \cite{messina1}.

\subsection{Trivalent Counterions.}
\begin{figure}
\hspace{-3cm}
   \epsfxsize=9cm
   \epsfysize=9cm 
~\hfill\epsfbox{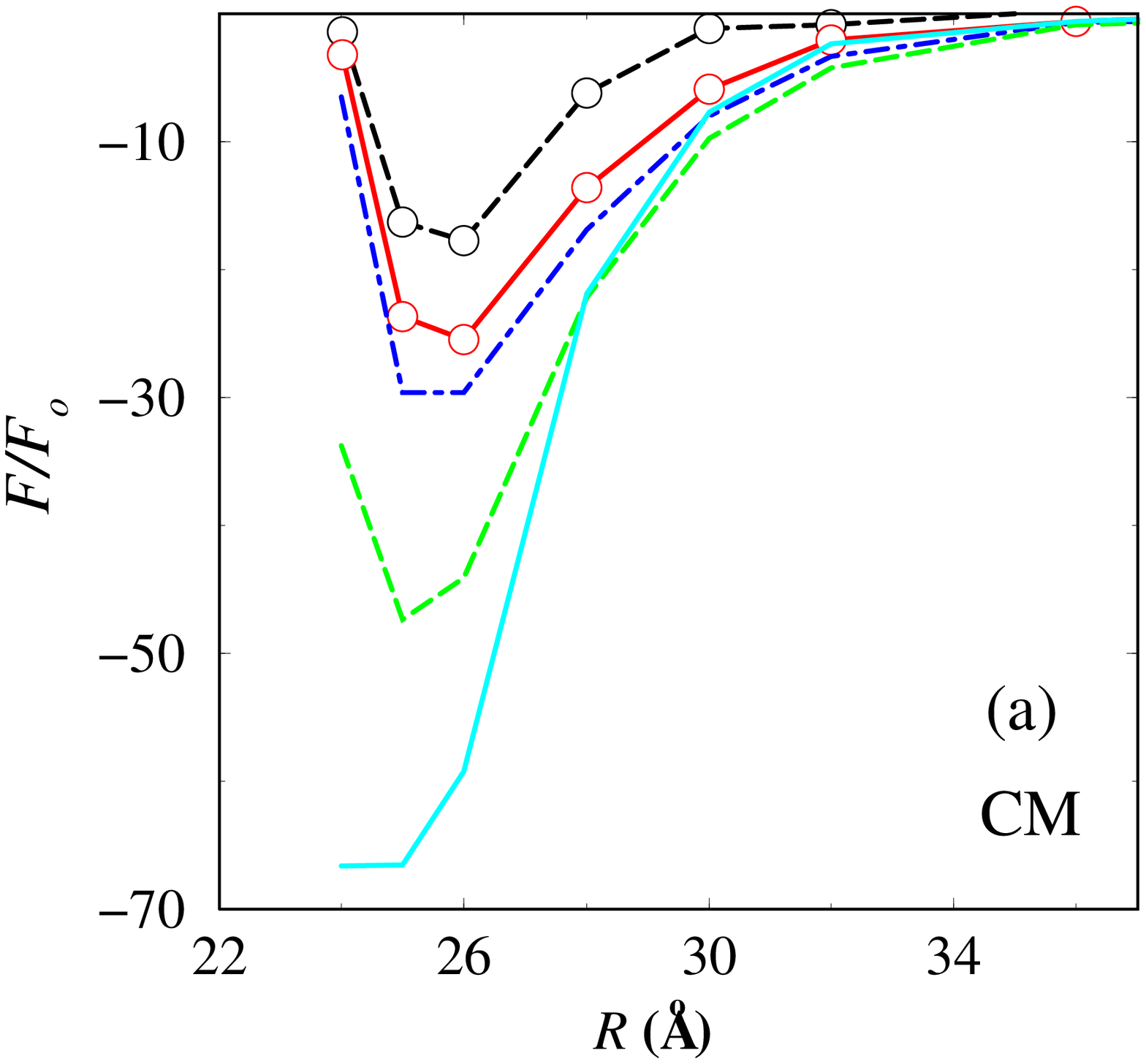} 
   \epsfxsize=9cm 
   \epsfysize=9cm 
~\hfill\epsfbox{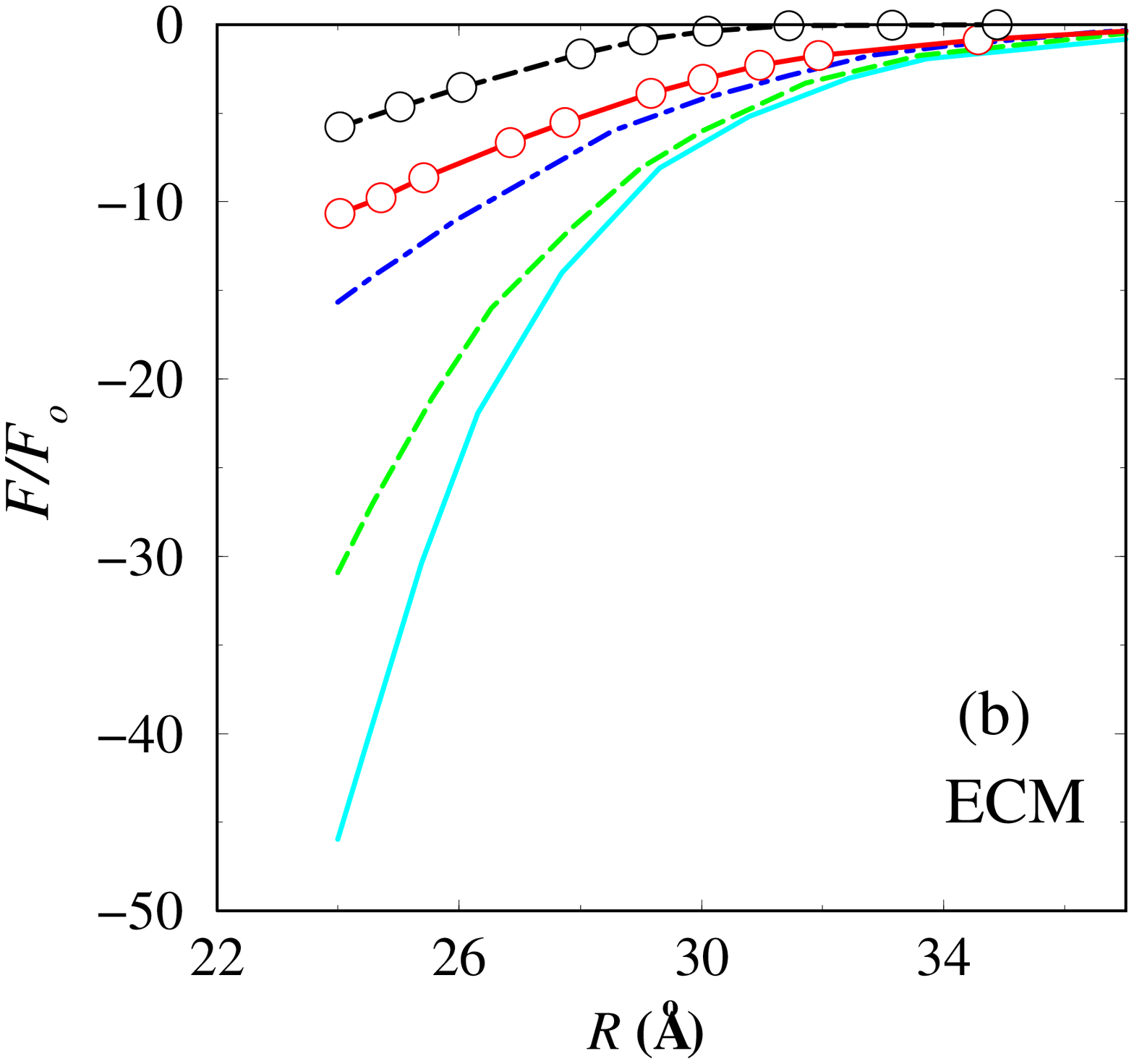}\hfill~
   \epsfxsize=9cm 
   \epsfysize=9cm 
~\hfill\epsfbox{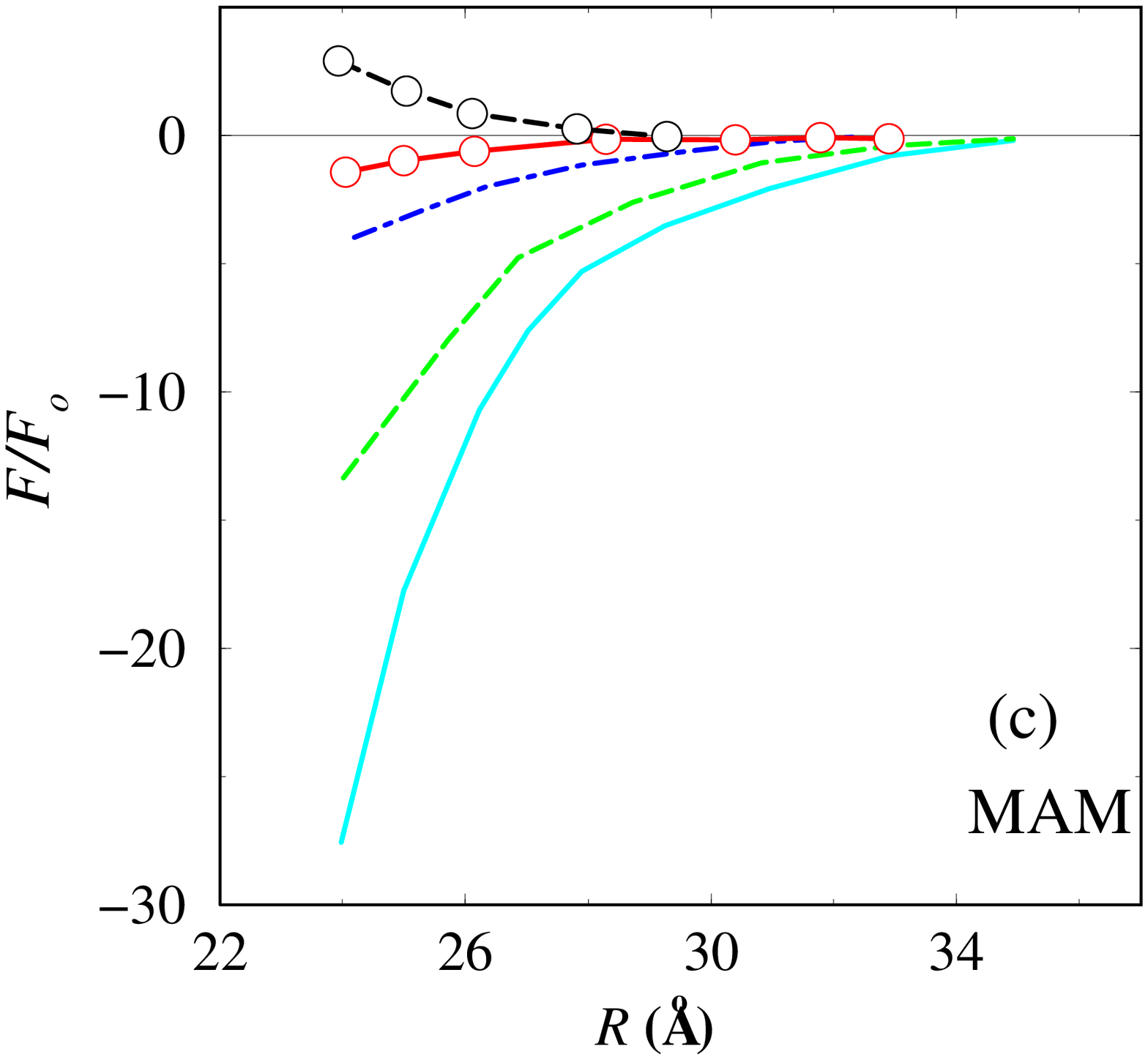}
\caption{Reduced DNA-DNA interaction force $F/F_0$ versus
  separation distance $R$ for a monovalent salt and trivalent
  counterions (parameter set 3 of Table I). The
  notation is the same as in Fig.~\ref{force_total_1_1}.}
 \label{force_total_3_1}
\end{figure}

Overwhelmingly attractive DNA-DNA interaction forces for trivalent
counterions and monovalent salt ions are plotted in
Fig.~\ref{force_total_3_1}. This attraction is stronger 
 than for the divalent counterions, in accordance with the
results of Ref.~\cite{gronbech}. Similar to the divalent counterion
case, the DNA-DNA attraction in the CM is  much stronger than
the attraction in the MAM for a given salt density. Evidently this is
related to the different counterion condensation patterns for different DNA
models \cite{oursecondDNApaper}. There is no salt-induced
stabilization for the ECM and MAM for the salt densities indicated in
Fig.~\ref{force_total_3_1}. Test runs with higher
salt concentrations (not shown here) reveal that the salt-induced stabilization in
the MAM does appear at $C_s=1.2$ mol/l when the repulsive interaction
force starts to descend towards zero. However, up to these high salt
concentrations, the attractive force in
the ECM steadily approaches zero from below. The same trend
holds in the CM as well.

\begin{figure}
\hspace{-3cm}
   \epsfxsize=9cm 
   \epsfysize=9cm 
~\hfill\epsfbox{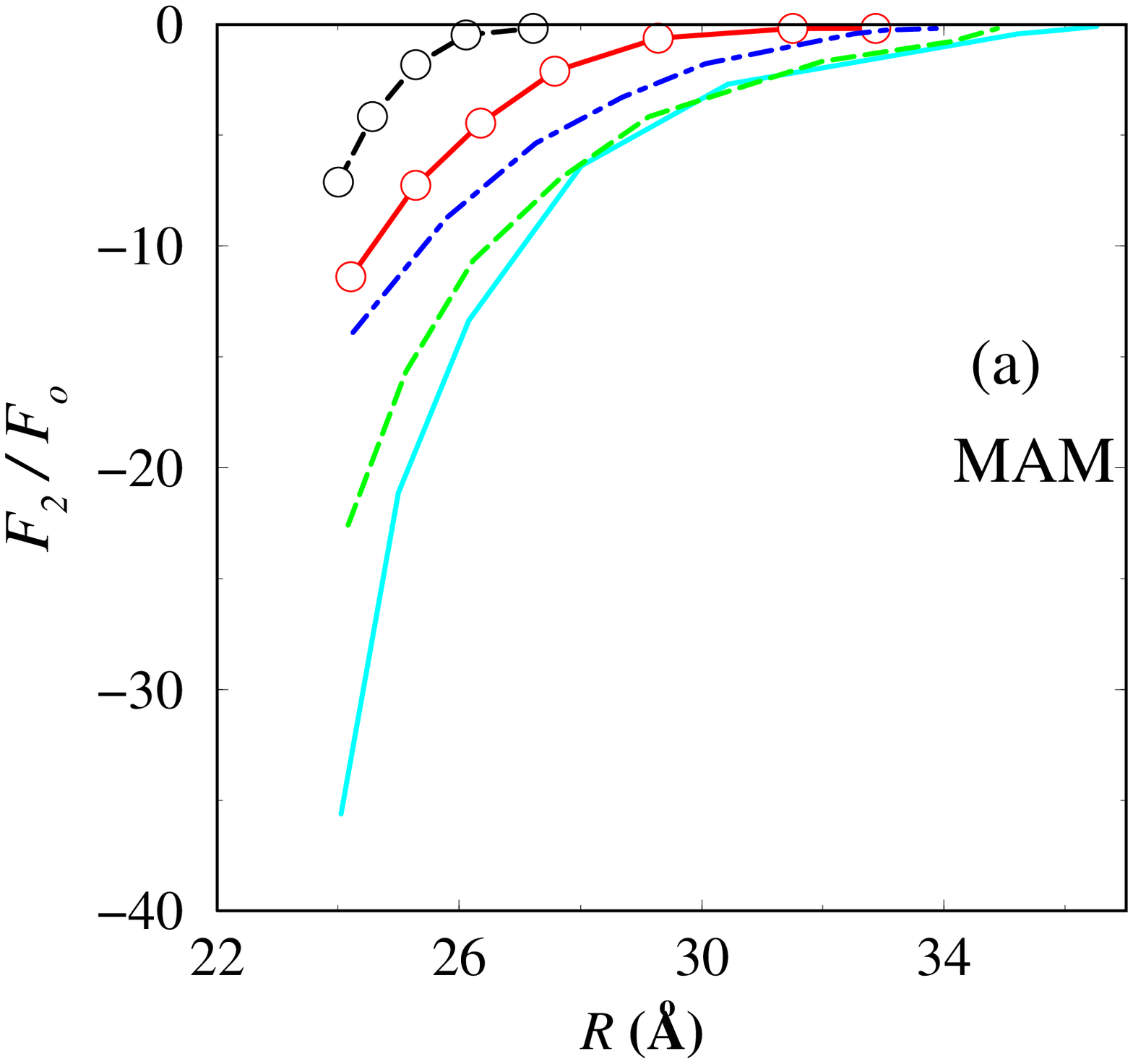}\hfill~ 
   \epsfxsize=9cm 
   \epsfysize=9cm 
~\hfill\epsfbox{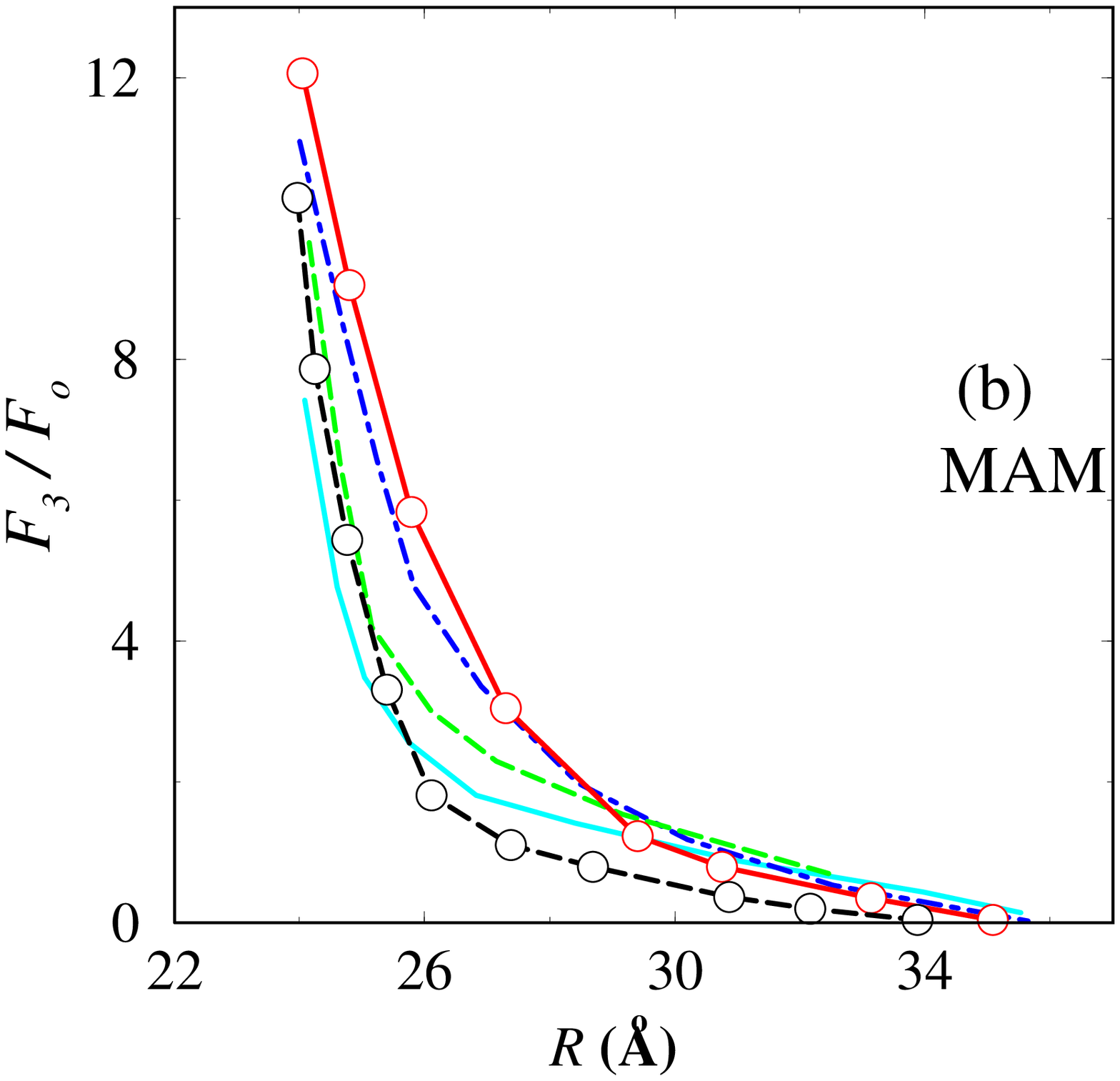}
\caption{(a) Reduced electrostatic  force $F_2/F_0$ and (b) entropic
  force $F_3/F_0$ components of the total
  interaction force $F(R)$ from the Fig.~\ref{force_total_3_1}c for the
  MAM. The notation is the same as in Fig.~\ref{force_total_1_1}.}
\label{force_mam_3_1}
\end{figure}

 The entropic force $F_3$ is non-monotonic against added
salt both in the ECM and MAM,  see Fig.~\ref{force_mam_3_1}b.   
 The reason is the following. The added
salt enhances the release of condensed trivalent counterions from the
DNA surface to the bulk \cite{burak2003}.
 However, the strong electrostatic correlations in the inner area
between the DNA molecules hinder this release. As a result, the
trivalent counterion release will be  asymmetric over the
 surface of DNA molecule. This will lead to an excess entropic
 force that pushes the DNA molecules away from each other. The 
osmotic pressure of added salt balances and at some salt density overcomes
 this entropic force, eventually driving the entropic force to zero.
 This entropic non-monotonicity
 does not survive in the total interaction force shown in
 Fig.~\ref{force_total_3_1}. This is because of the strong electrostatic attraction
 between the DNA molecules. Whereas the entropic repulsion is of
 the same order  for divalent and trivalent counterions, the
 electrostatic attraction for trivalent counterions is roughly
 twice as strong as for divalent counterions.     

\begin{figure}
\hspace{-3cm}
   \epsfxsize=9cm 
   \epsfysize=9cm 
~\hfill\epsfbox{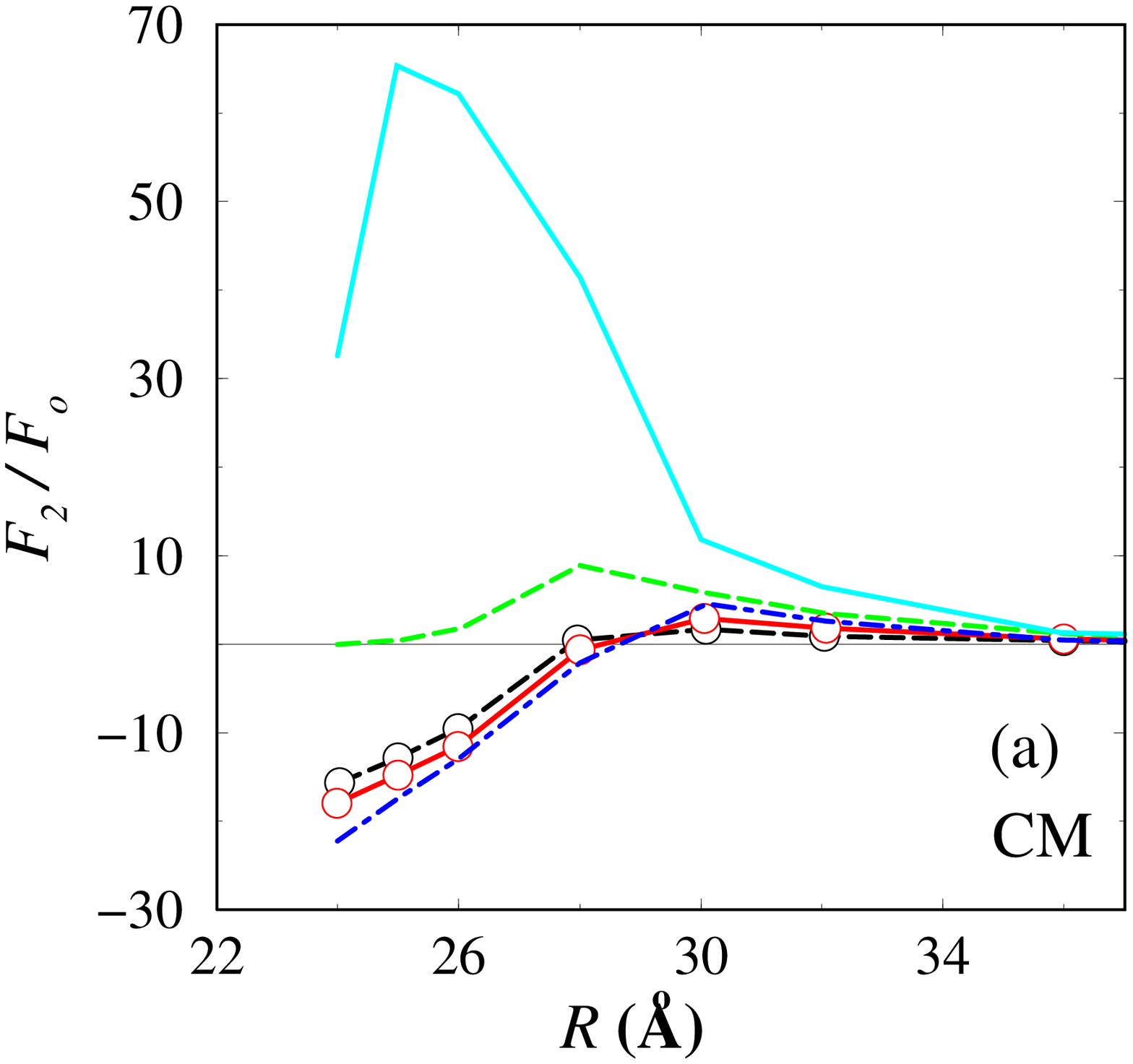}     
   \epsfxsize=9cm 
   \epsfysize=9cm 
~\hfill\epsfbox{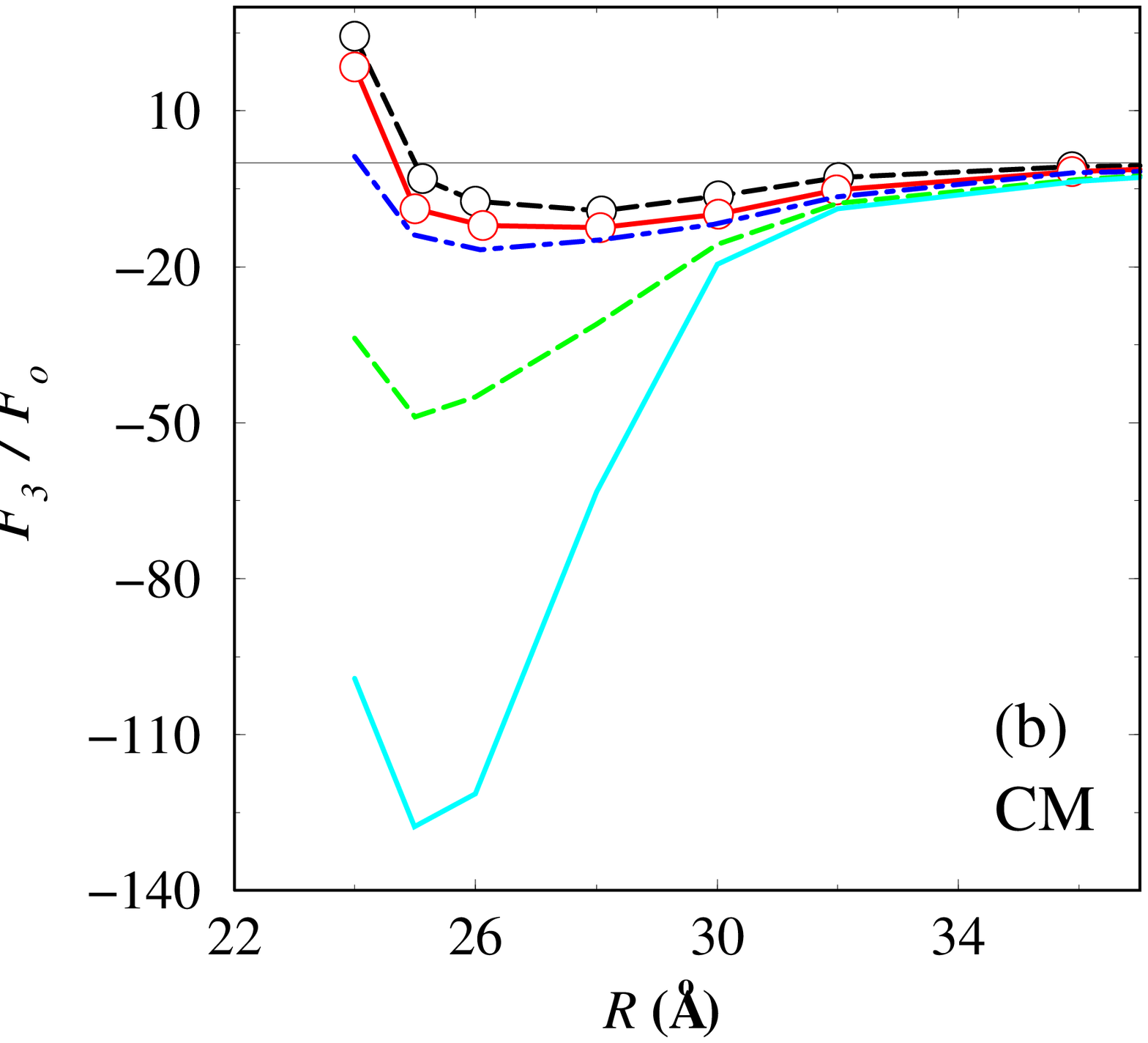}\hfill~
\caption{(a) Reduced electrostatic  force $F_2/F_0$ and (b) entropic
  force $F_3/F_0$ components of the total
  interaction force $F(R)$ from the Fig.~\ref{force_total_3_1}a for the
  CM. The notation is the same as in Fig.~\ref{force_total_1_1}.}
 \label{force_cm_3_1}
\end{figure}

The attraction in the CM for small
added salt densities $C_s$ in Fig.~\ref{force_cm_3_1} has again an
entropic origin \cite{allah}. For higher salt densities, however, the
attraction in the total interaction force at the short separation distances is
electrostatically driven. 

\section{simulation results for multivalent salt}
\subsection{Monovalent Counterions.}
The results obtained in  previous sections indicate that the
multivalent counterions generate
strong correlations inside the system and induce an
electrostatic attraction between the DNA rods in the monovalent salt
system. An intriguing
question is how this DNA-DNA attraction relates to DNA overcharging.
Going back to the single DNA
case,  considered in Ref.~\cite{oursecondDNApaper}, we observed no
overcharging for multivalent counterions and monovalent salt ions. Thus we conclude
that the overcharging effect is not a necessary condition for a DNA-DNA attraction to
take place. In other words, the electrostatic ion correlations, which
are not strong enough to induce the macroion overcharging, are able to
induce an attractive intermolecular force. On the other hand, in
Ref.~\cite{oursecondDNApaper} we have seen a DNA overcharging when
 multivalent salt ions were pumped into the  DNA suspension. 
When a DNA overcharging occurs, the ions form a sequence of
radial alternating charged layers around the DNA 
surface. The width of these layers is comparable
with the Debye screening length $\lambda_D$. 
The question we want to address here is to what extent the existence of such layers 
affects the DNA-DNA interaction.

\begin{figure}
\hspace{-3cm}
   \epsfxsize=9cm 
   \epsfysize=9cm 
~\hfill\epsfbox{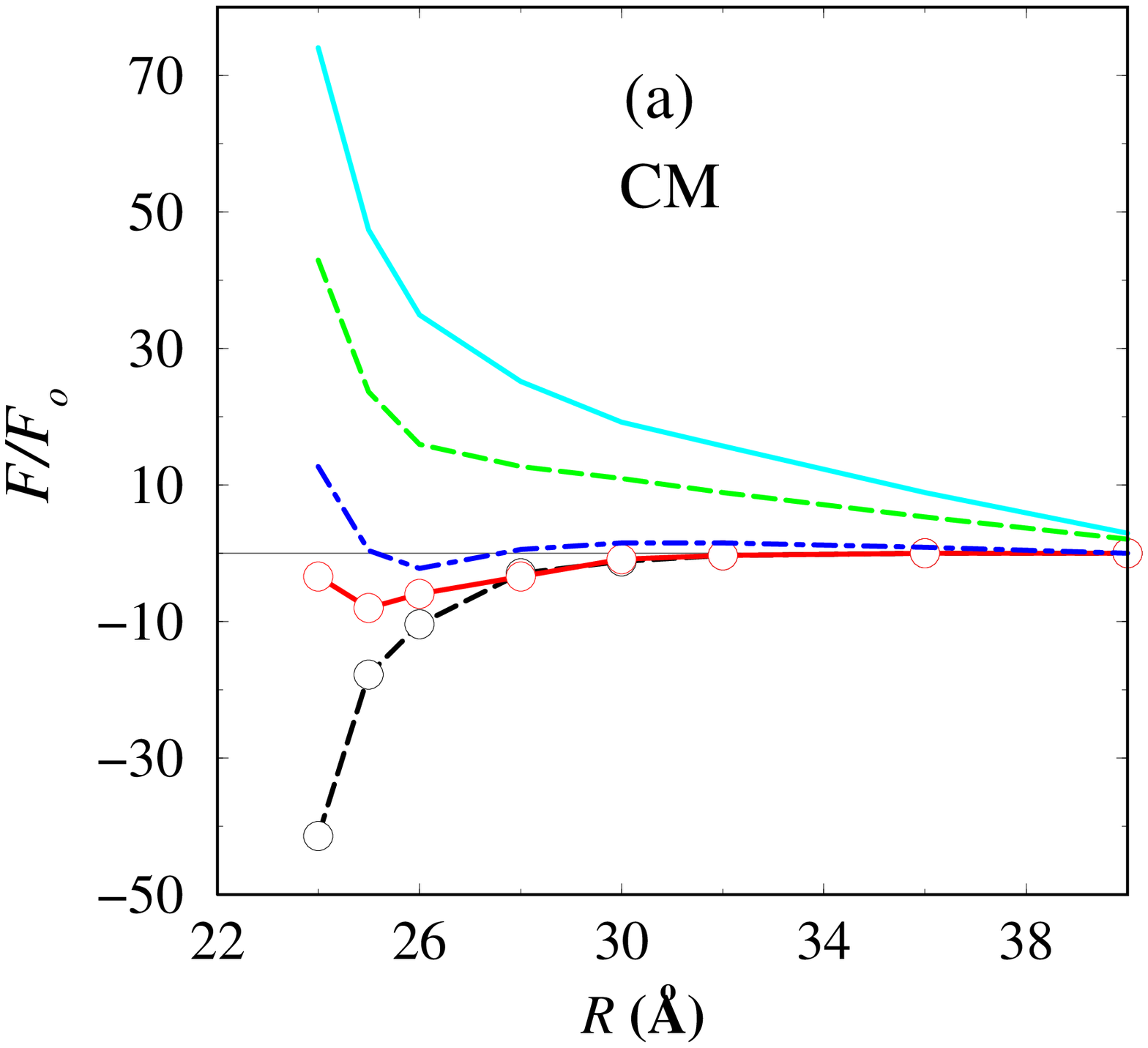} 
   \epsfxsize=9cm 
   \epsfysize=9cm 
~\hfill\epsfbox{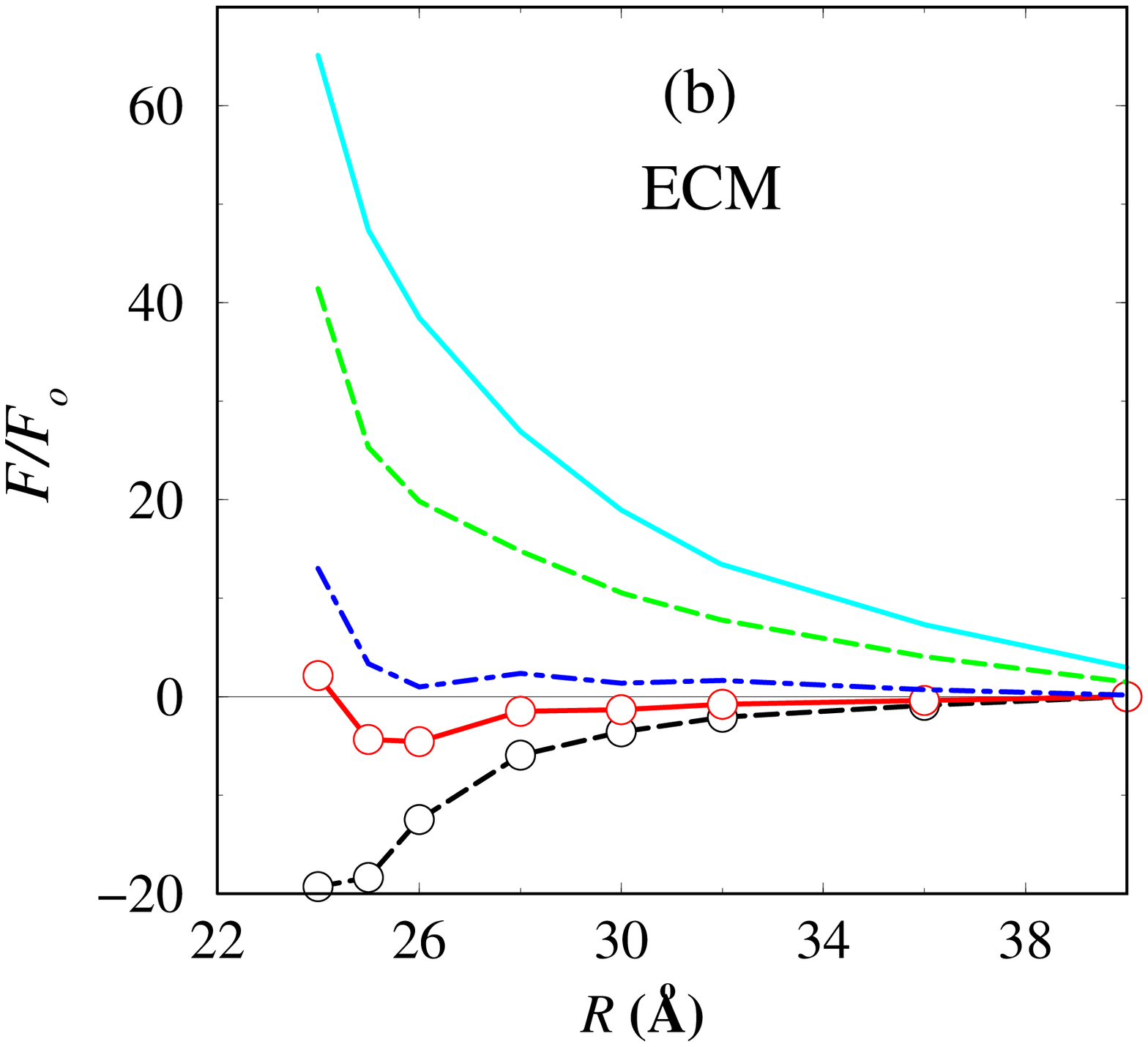}\hfill~
   \epsfxsize=9cm 
   \epsfysize=9cm 
~\hfill\epsfbox{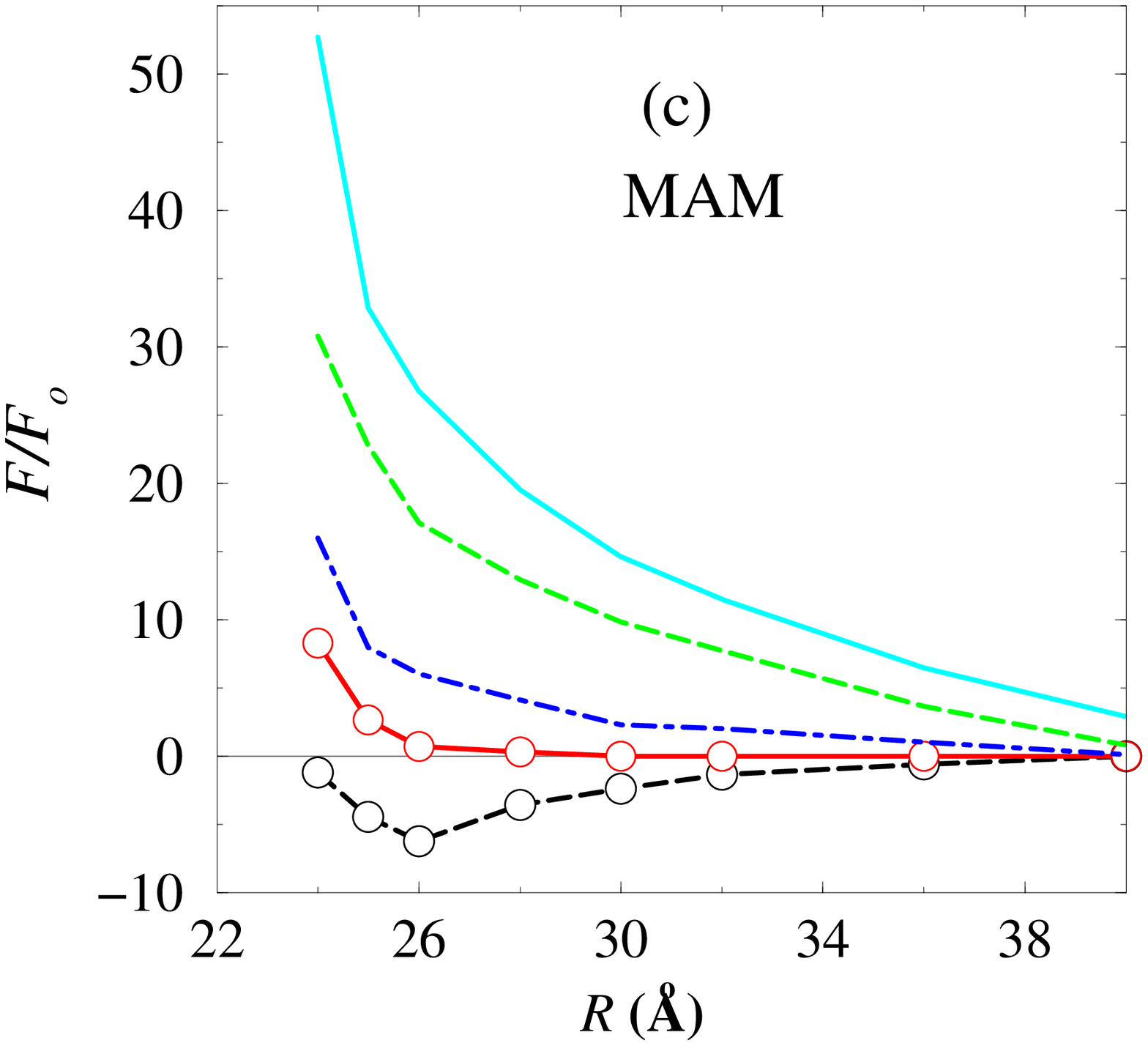}
\caption{Reduced DNA-DNA interaction force $F/F_0$ versus
  separation distance $R$ for a divalent salt and monovalent
  counterions (parameter set 4 of Table I). The
  notation is the same as in Fig.~\ref{force_total_1_1}.}
 \label{force_total_1_2}
\end{figure}

The total interaction forces for the divalent salt and monovalent
counterions are depicted in Fig.~\ref{force_total_1_2}. 
For all three DNA models, the DNA-DNA interaction is repulsive at small salt densities
 $C_s$ and transforms to an attraction for a sufficiently high $C_s$. 
 The DNA-DNA repulsion at lower salt is composed from the
 repulsive $F_2$ and $F_3$ components of the interaction force $F$. 
 In a similar way the attraction at a dense salt, which is strong for
 the CM  and weaker
for MAM, arises from both attractive electrostatic and entropic 
forces.
 For like-charged colloid particles at similar parameters for the small ions, an
 attraction was reported only for the electrostatic component of the
 interaction force \cite{wu1998}. 
A multivalent salt-induced precipitation
 of polyelectrolyte solution is also addressed in
 Ref.~\cite{belloni1995}. Whereas the van der Waals or hydrophobic
 forces could be presumed as a source for this attraction
 \cite{weiss2000cps,parsegian,sottas1999bj}, our 
 simulations are able to catch this effect without resorting to these forces.

 The DNA-DNA attraction at the divalent salt density $C_s$=0.71mol/l in
 Fig.~\ref{force_total_1_2}
 corresponds to the  overcharging of a single DNA molecule, see
 Ref.~\cite{oursecondDNApaper}.  
 Thus, an overcharging and entailed charged layers  near the DNA surface
 correlate with the DNA-DNA attraction in the divalent salt, at least when monovalent
 counterions are involved. This conclusion contradicts the results
 of Refs.~\cite{nguen_reentrant} where the onset of the DNA overcharging was
 considered to entail a repulsion between the DNA molecules, and
 therefore, a reentrance of the  DNA condensation. From our point of view, the
 discrepancy between our result and the result of Ref.~\cite{nguen_reentrant} is due
 to the different definition of overcharging. Whereas we count all
  charges near the DNA surface, only big multivalent ions were
 counted in Ref.~\cite{nguen_reentrant}. 

\subsection{Trivalent Counterions.}
\begin{figure}
   \epsfxsize=9cm 
   \epsfysize=9cm
~\hfill\epsfbox{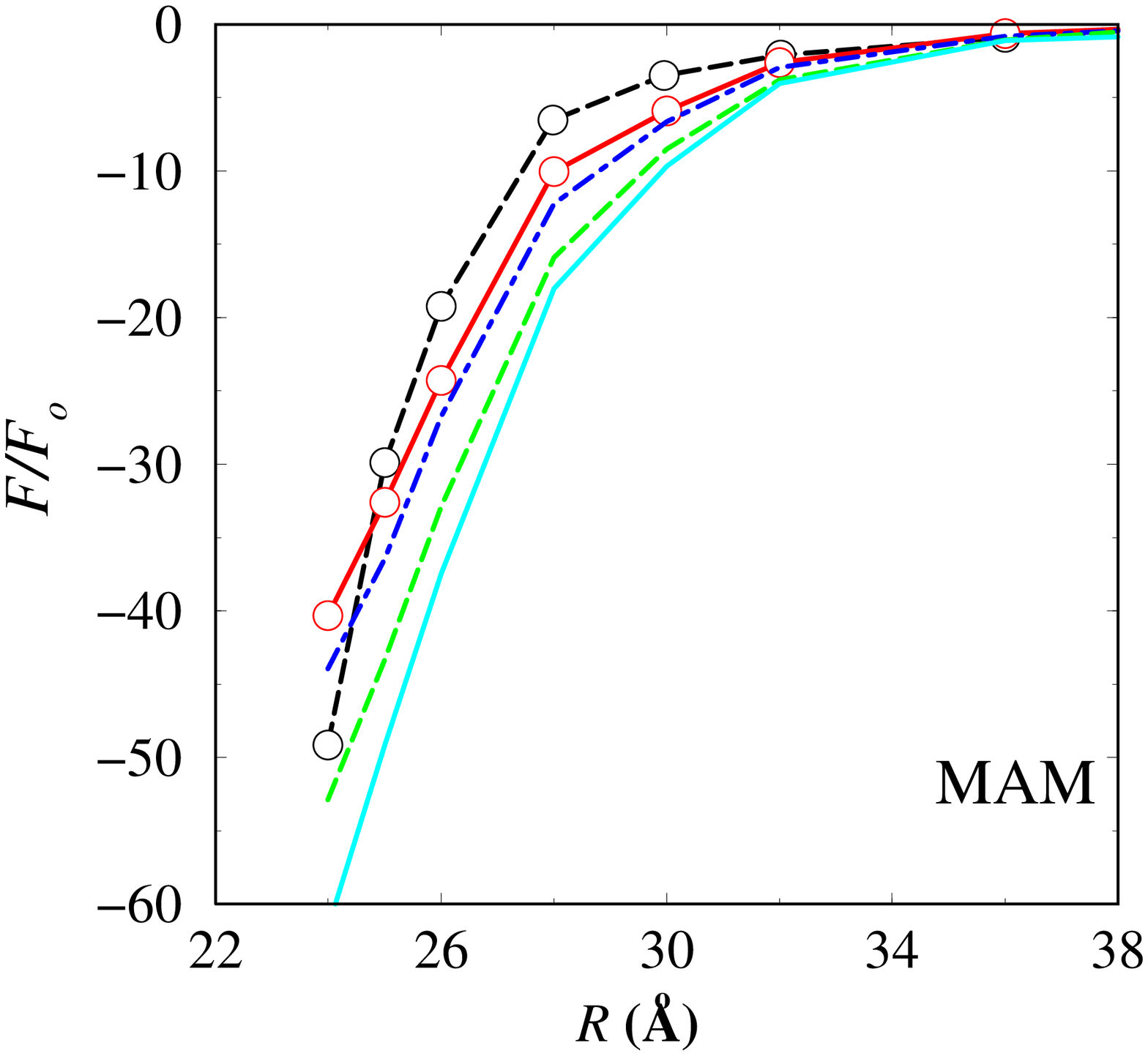}\hfill~
\caption{Reduced DNA-DNA interaction force $F/F_0$ versus
  separation distance $R$ for a divalent salt and trivalent
  counterions (parameter set 5 of Table I) and for the MAM. The
  notation is the same as in Fig.~\ref{force_total_1_1}.}
\label{force_ecm_3_2}
\end{figure}

 Simulation results for divalent salt and
 trivalent counterions are illustrated in Fig.~\ref{force_ecm_3_2}
 for the MAM. Now the DNA-DNA interaction force and both of its components $F_2$
 and $F_3$ (not shown here) are strongly attractive for all the calculated
 salt densities. The CM and ECM models exhibit
 similar trends. We note that at small salt densities, where no
 overcharging was found for a single DNA molecule
 \cite{oursecondDNApaper}, the obtained attractive force relates to
 the strong charge correlations in system. Thus the claim that the overcharging
 effect is a sole factor that induces the DNA attraction, is
 not correct. Broadly speaking, there is a competition between the correlations that
 induce an attraction between the DNA molecules (multivalent counterion
 induced correlations) and the overcharging effect, which
 induce a DNA-DNA attraction (multivalent salt-induced
 correlations). To understand the physics of this
 competition we have analyzed the tendencies of these two 
 correlation effects against the increase of 
 the salt density. Fig.~\ref{force_ecm_3_2} shows that the
 interaction forces decay monotonically with increasing salt concentration $C_s$.
 This trend is in contrast to the results for a divalent salt and
 monovalent counterions in Fig.~\ref{force_total_1_2}, where more
 salt-induced more attraction. Thus, the main contribution to the DNA-DNA
attraction comes from the strong correlations between the two
strongly correlated layers of trivalent counterions on the DNA surfaces.  Pumping more
divalent salt into the system destroys the two-dimensional crystal structure and
correlations. But the interaction force remains attractive,
 mainly due to the additional overcharging of the DNA. As a result,
the DNA-DNA attraction survives for a dense salt, in opposite to the
case of trivalent counterions and monovalent salt 
shown in Fig.~\ref{force_total_3_1}.   

\section{conclusion}
We have studied the interaction forces between a pair of DNA molecules in
an electrolyte that contains a mixture of monovalent and  multivalent
ions. Three models for the DNA shape, employed in our simulations, indicate the
importance of the DNA geometry on the  electrostatic and entropic forces
in the DNA conformations. We show that the DNA-DNA attraction is
related to the charge correlations in strongly charged systems. We
distinguish between multivalent counterion and  multivalent salt
induced attractions. In general, the higher the mean valency of the
microions in the solution, the stronger is the mutual attraction.
Below we shortly summarize the main results of this manuscript.

For the multivalent counterions, the DNA shape is an essential
contributor to the interaction forces. Thus:\\
i) For no added salt the DNA-DNA attraction in the CM is related
to the Coulomb  depletion mechanism. This depletion effect results in
an attractive entropic force. However such an attraction mechanism does
not exist in the MAM. \\
ii) For the nonzero added salt cases, an attraction in the CM is mainly
due to a combination of electrostatic and entropic forces. However the
attractive force in the MAM always has an electrostatic origin. The
entropic force in the MAM is always repulsive.\\
iii) There is a non-monotonic force-salt dependence at a fixed separation
distance. For divalent counterions, there is a change
of the interaction force from the repulsion to an attraction, and then back
to zero, which we call salt-induced stabilization.\\
iv) An increase of the salt concentration suppresses the charge correlations
and thus effectively screens the  DNA-DNA attraction.

For the multivalent added salt, the DNA shape has a minor effect on
the interaction forces. DNA-DNA interaction forces are stronger for the
CM than  for the MAM.  Further trends are:\\
i) An increase of the divalent salt density at a fixed monovalent ion
number drives the DNA-DNA interaction 
force towards attraction. Both  DNA molecules are overcharged in the
attractive force regime. \\
ii) For trivalent counterions the addition of divalent salt decreases
the DNA-DNA attraction. The more the DNA becomes overcharged, the less
is the attraction between the DNA molecules.\\ 
iii) The correlation effects related to multivalent counterions have
a greater influence on the DNA attraction than the correlation effects
related to multivalent salt ions. In other words, an
overcharging-induced attraction is weak compared to the
counterion-induced attraction.  
 
We would like to make some comments about the range of the DNA-DNA
attraction which directly influences the phase
diagram of DNA solutions \cite{gast1983,lekker1992}. Compared to the Debye screening
length, the attraction forces between
the DNA molecules are long-ranged (they are beyond the screening
length). In contrast, for colloids, usually the attraction is short ranged in comparison with
the screening length \cite{larsen1997nature}.
Thus, the calculated attractive DNA-DNA forces can lead to phase separation in
 DNA solutions but they are not the dominant driving forces of DNA condensation:
The actual ``arbiters'' of phase instability in
macroion solutions seem to be many-body interactions
\cite{potemkin2002,ha1997nonadditivity,knott2001PBattraction}. 
 These forces separate the
 DNA assemble into DNA rich and DNA 
poor regions even for a purely repulsive
pairwise interaction \cite{roij2000,warren2000jcp,dan1996bj}.
Another intriguing effect is the appearance of a cholesteric phase of
DNA condensates, which arises from DNA chirality
and  many-body interactions \cite{deng1999,chirality,strzelecka1987and1988,livolant}. 

Finally, let us comment on some limitations of our model.
Our choice of a continuum dielectric
model has the intention to separate the purely
electrostatic effects from the effects of hydration and the molecular
structure of the solvent.  
In many applications,
including strong polyelectrolytes and high salt concentrations,
continuum dielectric models (the primitive electrolyte model) have
been successful (see
Refs.~\cite{simulationswithmolecularwater,lyubartsev1995pre}).
However, strictly speaking, the continuum model is not
justified at small ion-DNA and ion-ion separations where the molecular
nature of the solvent is no longer negligible
\cite{gonzales-mozuelos2000,belloni1997,kusalik1988,allahyarovsolventeffect}.
 At these distances the effective  
(mean-force) potentials of ions have one to two oscillations
around the potential of primitive model
 \cite{gulbrand1989,kjellander2001solvent,kusalik1988,allahyarovsolventeffect,pettitt1986jcp,LIE}.

It is worth to mention the  complete neglect of the specific binding
 (or chemisorbtion) of counterions to the DNA grooves in our
 simulations. Active binding \cite{chiu2000jmb} can be taken into
 account through the incorporation  of a full water description 
\cite{korolev,korolev2001biopolymers,korolev2002biophys,young2} or via the
 implementation an additional sticky potential to certain ions in
 parts of the DNA areas. A cylindrical well around the DNA due to the solvent mediated
mean-field potential or specific short-range ion-DNA interactions, can
 also replace the specific "bonding" of ions to the DNA surface. 

Other effects not accounted for in the dielectric continuum model are   
the dielectric discontinuity and dielectric saturation effects.
The former effect emerges due to the polarization of the DNA surface
and affects the ion distribution outside the DNA core \cite{linse1986}. 
It rapidly drops off for large distances from the surface \cite{gavryushov}. 
The latter effect is related to the water anisotropy near the DNA surface
 \cite{saenger1987} and can be accounted for through a
distance-dependent $\epsilon$ in the electrostatic potentials
\cite{lyubartsev1995,yang1995bj,jaya2,jayaram1998jpc,maarel1999,mazur,petsev2000bj,luka1,jayaram1996epsilonR}. A decade ago there was a widely accepted
perception that the attraction between the DNA molecules cannot be
explained by the electrostatic forces
 \cite{parsegian,hydration_forces,petsev2000}. However the 
theoretical investigations and 
simulations of strongly correlated systems 
 do not support this claim \cite{bloomfield1997,lyubartsev1995}. 
Finally, we neglect the salt-induced decrease of the DNA rigidity
in salted solutions
\cite{ariel2003,duguid1995,ibragimova1998,liu2000jcp,williams,lu2002,baumann1997}
and  dielectric permeability of the solvent \cite{kusalik1988}.
To include all these additional complications into a simulation
remains a very challenging task for the future. 

\acknowledgments
E. A. thanks A. Cherstvy, R. Podgornik and A. Parsegian for fruitful
discussions of some results of the current paper.

\begin{table}
\caption{Parameter sets used for the simulations of DNA-DNA interactions.}
\begin{tabular}{lcc}
Set & $q_c$ & $q_s$  \\ 
\tableline
  1 &   1  &  1  \\
  2 &   2  &  1  \\
  3 &   3  &  1  \\ 
  4 &   1  &  2  \\ 
  5 &   3  &  2  \\ 
\end{tabular} 
\end{table}

\newpage
\newpage

\newpage

\newpage

\newpage

\newpage

\newpage

\newpage

\newpage

\newpage

\newpage

\newpage

\newpage

\begin{references} 
\bibitem{pelta1996} J. Pelta, F. Livolant, J.-L. Sikorav,
J. Biological Chemistry {\bf 271}, 5656 (1996).
\bibitem{bloomfield1991biopolymers} V. A. Bloomfield, Biopolymers {\bf
    31}, 1471 (1991).
\bibitem{bloomfield1997} V. A. Bloomfield, Biopolymers {\bf 44}, 269 (1997). 
\bibitem{bloomfield1996cosb} V. A. Bloomfield,
  Curr. Opin. Struct. Biol. {\bf 6}, 334 (1996).
\bibitem{shui1} X. Shui, L. McFail-lsom, G. G. Hu, L. D. Williams,
  Biochemistry {\bf 37}, 8341 (1998).
\bibitem{tajmir} H. A. Tajmir-Riahi, M. Naoui. R. Ahmad,
  J. Biomol. Struct. Dyn. {\bf 11}, 83 (1993); I. Fita, J. L. Campos,
  L. C. Puigjaner, J. A. Subriana, J. Mol. Biol. {\bf 167}, 157
  (1983).
\bibitem{wierenga1996} A. M. Wierenga, A. P. Philipse, J. Colloid
  interface Sci. {\bf 180}, 360 (1996).
\bibitem{kjellander1988colloidintsci} R. Kjellander, S. Mar\v celja,
    J. P. Quirk, J. Colloid. Interface Sci. {\bf 126}, 194 (1988).
\bibitem{kepler} G. M. Kepler, S. Fraden, Phys. Rev. Letters {\bf 73}, 356 (1994).
\bibitem{crocer1996prl} J. C. Crocker, D. G. Grier, Phys. Rev. Lett.
  {\bf 77}, 1897 (1996). 
\bibitem{larsen1997nature} A. E. Larsen, D. G. Grier, Nature
  {\bf 385}, 231 (1997). 
\bibitem{wennerstrom1991acis} H. Wennerstr{\"o}m, A. Khan, B. Lindman,
  Adv. Colloid. Interface Sci. {\bf 34}, 433 (1991). 
\bibitem{kekicheff} P. Kekicheff, S. Mar\v celja, T. J. Senden, V. E. Shubin,
  J. Chem. Phys. {\bf 99}, 6098 (1993).
\bibitem{saenger} W. Saenger, 1984 Principles of Nucleic Acid
  structure. Springer-Verlag, New-York.
\bibitem{benbasat1984} J. A. Benbasat,  Biochemistry {\bf 23}, 3609 (1984).
\bibitem{deng1999} H. Deng, V. A. Bloomfiled, Biophysical J. {\bf 77},
  1556 (1999).
\bibitem{Solis1999} F. J. Solis, M. O. de la Cruz, Phys. Rev. E {\bf
  60}, 4496 (1999). 
\bibitem{lamm1993} G. Lamm, G. R. Pack, Int. J. Quant. Chem. {\bf 20}, 213
(1993); Biopolymers {\bf 34},  227 (1994). 
\bibitem{nilsson} L. G. Nilsson, L. Gulbrand, L. Nordenski\"old,
  Mol. Phys. {\bf 72}, 177 (1991).
\bibitem{lyubartsev1995} A. P. Lyubartsev, L. Nordenski\"old, J. Phys. Chem. {\bf
  99}, 10373 (1995).
\bibitem{Deng} H. Deng, V. A. Bloomfeld, J. M. Benevides, G. J. Thomas Jr, Nucleic acids
research {\bf 28}, 3379 (2000).
\bibitem{khan} M. O. Khan, B. J{\"o}nsson, Biopolymers {\bf 49}, 121 (1999).
\bibitem{ha1998attraction} B.-Y. Ha, A. J. Liu,  Phys. Rev. Letters
  {\bf 81}, 1011 (1998).
\bibitem{deserno2002attraction}
M. Deserno, A. Arnold, Ch. Holm, Macromolecules, American Chemical Society,
Preprint (2002).
\bibitem{guldbrand} L. Guldbrand, B. J\"onsson, H. Wennerstr\"om,
  P. Linse, J. Chem. Phys. {\bf 80}, 2221 (1984); L. Guldbrand,
  L. G. Nilsson, L. Nordenski\"old J. Chem. Phys. {\bf 85}, 6686 (1986). 
\bibitem{bolhius} P. G. Bolhuis, T. \AA kesson, B. J\"onsson,
  J. Chem. Phys. {\bf 98}, 8096 (1993).
\bibitem{cai2002} X. E. Cai, J. Yang, Biophys. J. {\bf 82}, 357 (2002).
\bibitem{vervey} E. J. W. Vervey, J. T. G. Overbeek, Theory of stability of
  Lyophobic Colloids, Elsevier, Amsterdam (1948).
\bibitem{reiner1993acis} E. S. Reiner, C. J. Radke, Adv. Colloid
  Interface Sci {bf 47}, 59 (1993).
\bibitem{mills} P. Mills, C. F. Anderson, M. T. Record, J. Phys. Chem. {\bf
  89} 3984 (1985).
\bibitem{marcus1955} R. A. Marcus, J. Chem. Phys. {\bf 23}, 1057 (1955).
\bibitem{fogo} F. Fogolari, P. Zuccato, G. Esposito, P. Viglino,
  Biophysical J. {\bf 76}, 1 (1999).
\bibitem{lin} P. J. Lin-Chung, A. K. Rajagopal, Phys. Rev. E {\bf 52}, 901
  (1995).
\bibitem{gilson} M. K. Gilson, M. E. Davis, B. A. Luty,
  J. A. McCammon, J. Phys. Chem. {\bf 97}, 3591 (1993).
\bibitem{wagner} K. Wagner, E. Keyes, T. W. Kephart, G. Edwards, Biophysical J.
  {\bf 73}, 21 (1997).
\bibitem{vlachy1999} V. Vlachy, Ann. Rev. Phys. Chem.  {\bf 50}, 145 (1999).
\bibitem{gavryushov} S. Gavryushov, P. Zielenkiewicz, Biophysical
  J. {\bf 75}, 2732 (1998).  
\bibitem{rescic1997} J. Rescic, V. Vlachy, L. B. Bhuiyan, C. W. Outhwaite, J. Chem. 
Phys. {\bf 107}, 3611 (1997).
\bibitem{das1995modifiedPB} T. Das, D. Bratko, L. B. Bhuiyan,
  C. W. Outhwaite,  J. Chem. Phys {\bf 99}, 410 (1995).  
\bibitem{Netz1}  R. R. Netz, H. Orland, Euro. Phys. J. E  {\bf 1},
  203 (2000). 
\bibitem{sood1991} A. K. Sood, Solid State Physics {\bf 45}, 1 (1991).
\bibitem{coen1995} C. J. Coen, H. W. Blanch, J. M. Prausnitz, AICHE J.
  {\bf 41} 996 (1995).
\bibitem{vlachy1993} V. Vlachy, H. W. Blanch, J. M. Prausnitz, AICHE
  J. {\bf 39},  215 (1993).
\bibitem{farnum1999bj} M. Farnum, C. Zukoski, Biophysical J. {\bf 76}, 2716 (1999).
\bibitem{spalla2000hamaker} O. Spalla, Curr. Opinion Colloid. Interface
Sci. {\bf 5}, 5 (2000).
\bibitem{weiss2000cps} A. Weiss, M. Ballauff,
  Colloid. Polym. Sci. {\bf 278}, 1119 (2000).
\bibitem{hnc_attraction}
R. Kjellander, S. Mar\v celja, Chem. Phys. Lett. {\bf 49}, 112 (1984);
C. L. Steven, T. M. Glenn,  Adv. Chem. Phys. {\bf 56}, 141 (1984);
S. E. Feller, D. A. McQuarrie, Mol. Phys. {\bf 80}, 721 (1993).
\bibitem{pincus} N. Gr{\o}nbech-Jensen, K. M. Beardmore, P. Pincus,
Physica A {\bf 261}, 74 (1998); A. W. C. Lau, P. Pincus et al, Phys. Rev. E {\bf
  63}, 051604-1 (2001); P. Pincus, S. A. Safran, Euro. Phys. Letters
{\bf 42}, 103 (1998). 
\bibitem{harau2002jcp} L. Harnau, J.-P. Hansen, J. Chem. Phys. {\bf 116}, 9051 (2002).
\bibitem{kjellander1986}  R. Kjellander, S. Mar\v celja,
    J. Phys. Chem. {\bf 90}, 1230 (1986). 
\bibitem{attard1988} P. Attard, D. J. Mitchell, B. W. Ninham,
  J. Chem. Phys. {\bf 88}, 4987 (1988);  P. Attard, D. J. Mitchell,
  B. W. Ninham, J. Chem. Phys. {\bf 89}, 4358 (1988).  
\bibitem{podgornik1990} R. Podgornik, J. Phys. A {\bf 23}, 275
  (1990). 
\bibitem{stevens1990} M. J. Stevens, M. O. Robbins,
  Europhys. Lett. {\bf 12}, 81 (1990). 
\bibitem{lau2000prl} A. W. C. Lau, D. Levine, P. Pincus,
  Phys. Rev. Lett. {\bf 84}, 4116 (2000).
\bibitem{roij2000} R. van Roij, J. Phys. Cond. Matter {\bf 12},
  A263 (2000).
\bibitem{roij1997and1999} R. van Roij, J.-P. Hansen, Phys. Rev. Lett. {\bf
  79}, 3082 (1997); R. van Roij, M. Dijkstra, J.-P. Hansen,
  Phys. Rev. E {\bf 59}, 2010 (1999).  
\bibitem{Spalla} O. Spalla, L. Belloni, J. Chem. Phys. {\bf 95}, 7689 (1991).
L. Belloni, O. Spalla, jchem.p 107 (1997) 465
\bibitem{hribar1997} B. Hribar, V. Vlachy, J. Phys. Chem. {\bf 101}, 3457 (1997).
\bibitem{netz1999epland1999preand2001epje} 
R. R. Netz, H. Orland, Europhys. Lett.  {\bf 45}, 726 (1999); 
R. R. Netz, Eur. Phys. J. E {\bf 5}, 557 (2001);
R. R. Netz, Phys. Rev. E {\bf 60}, 3174 (1999);
A. G. Moreira, R. R. Netz, Phys, Rev. Letters {\bf 87}, 078301 (2001).
\bibitem{goulding1999epl} D. Goulding, J.-P. Hansen, Europhys. Letters
  {\bf 46}, 407 (1999).
\bibitem{ariel2003} G. Ariel, D. Andelman, Phys. Rev. E {\bf 67}, 011805 (2003).
\bibitem{burak2003} Y. Burak, G. Ariel, D. Andelman, submitted to Biophys J. 2003.
\bibitem{gronbech} N. Gr{\o}nbech-Jensen, R. J. Mashl, R. F. Bruinsma,
  W. M. Gelbart, Phys. Rev. Letters {\bf 78}, 2477 (1997);
  N. Gr{\o}nbech-Jensen, K. M. Beardmore, Physica A {\bf 261}, 74 (1998).
\bibitem{levin} Y. Levin, J. J. Arenzon, J. F. Stilck, Phys. Rev. Letters {\bf
  83}, 2680 (1999); J. J. Arenzon, J. F. Stilck, Y. Levin,
Eur. Phys. J. B {\bf 12}, 79 (1999). 
\bibitem{delrow} J. J. Delrow, J. A. Gebe, J. M. Schurr, Inc. Biopoly {\bf
    42}, 455 (1997).
\bibitem{Linse1999} P. Linse, V. Lobaskin, Phys. Rev. Lett. {\bf 83}, 4208 (1999); 
J. Chem. Phys. {\bf 112}, 3917 (2000).
\bibitem{kjellander1992chempot} R. Kjellander, T. Akesson ,
  B. J\"onsson,S. Mar\v celja, J. Chem. Phys. {\bf 97}, 1424 (1992).
\bibitem{rouzina} I. Rouzina, V. A. Bloomfield, J. Chem. Phys. {\bf 100},
  9977 (1996).
\bibitem{lukatsky1999fluctuation} D. B. Lukatsky, S. A. Safran,
  Phys. Rev. E {\bf 60}, 5848 (1999).
\bibitem{murthy} C. S. Murthy, R. J. Bacquet, P. J. Rossky,
  J. Phys. Chem. {\bf 89}, 701 (1985).
\bibitem{vlachy} V. Vlachy, A. D. J. Haymet, J. Chem. Phys. {\bf 84}, 5874 (1986).
\bibitem{paulsen1988} M. D. Paulsen, C. F. Anderson, M. T. Record, Biopolymers
  {\bf 27}, 1249 (1988).
\bibitem{mills1986} P. Mills, M. D. Paulsen, C. F. Anderson,
  M. T. Record Jr., Chem. Phys. Letters {\bf 129}, 155 (1986).
\bibitem{paulsen1987} M. D. Paulsen, B. Richey, C. F. Anderson,
  M. T. Record Jr., Chem. Phys. Letters {\bf 139}, 448 (1987).
\bibitem{katchalsky} A. Katchalsky, Pure Appl. Chem. {\bf 26}, 327 (1971).
\bibitem{jaya} B. Jayaram, D. L. Beveridge,
  Annu. Rev. Biophys. Biomol. Struct. {\bf 25}, 367 (1996).
\bibitem{yang1995bj} L. Yang, A. Weerasinghe, B. M. Pettitt, Biophys
  J. {\bf 69}, 1519 (1995).
\bibitem{ourfirstDNApaper} E. Allahyarov, H. L{\"o}wen,  Phys. Rev. E
  {\bf 62}, 5542 (2000). 
\bibitem{oursecondDNApaper} E. Allahyarov, H. L{\"o}wen, G. Gompper,
  accepted in Phys. Rev. E (2003). 
\bibitem{allah} E. Allahyarov, I. D'Amico, H. L\"owen,
  Phys. Rev. Letters {\bf 81}, 1334 (1998).
\bibitem{parsegian} D. C. Rau, B. Lee, V. A. Parsegian,
  Proc. Natl. Acad. Sci {\bf 81}, 2621 (1984); R. Podgornik, D. C. Rau,
  V. A. Parsegian, Biophys. J. {\bf 66}, 962 (1994); R. Podgornik,
  D. C. Rau,
  V. A. Parsegian, Macromolecules {\bf 22}, 1780 (1989); H. H. Strey,
  V. A. Parsegian, R. Podgornik, Phys. Rev. Letters {\bf 78}, 895 (1997); R. Podgornik,
  H. H. Strey, K. Gawrisch, D. C. Rau, A. Rupprecht, V. A. Parsegian,
  Proc. Nat. Acad. Sci. USA {\bf 93}, 4261 (1996); S. Leikin,
  V. A. Parsegian, D. C. Rau, R. P. Rand, Annu. Rev. Phys. Chem. {\bf 44}, 369 (1993). 
\bibitem{schneider1998} B. Schneider, K. Patel, H. M. Berman,
  Biophys. J. {\bf 75}, 2422 (1998).
\bibitem{gonzales-mozuelos2000}  P. Gonz\'alez-Mozuelos, N. Bagatella-Flores,  Physica
  A {\bf 286}, 56 (2000). 
\bibitem{barrat1996} J. L. Barrat, J. F. Joanny, Adv. Chem. Phys. {\bf
    94}, 1 (1996).   
\bibitem{oosawa1971} F. Oosawa, {\it Polyelectrolytes} (Marcel Dekker,
    New York, 1971).
\bibitem{bloomfield1999enzymol} V. A. Bloomfield, I. Rouzina, Methods
    Enzymol. {\bf 295}, 364 (1999).
\bibitem{hecht1995} J. L. Hecht, B. Honig, Y. K. Shin, W. L. Hubbell,
  J. Phys. Chem. {\bf 99}, 7782 (1995).
\bibitem{kor1} A. A. Kornyshev, S. Leikin, J. Chem. Phys. {\bf 107}, 3656 (1997);
 Proc. Natl. Acad. Sci. USA {\bf 95}, 13579 (1998);
 Biophys. J. {\bf 75}, 2513 (1998); J. Chem. Phys. {\bf 108}, 7035(E)
 (1998); A. G. Cherstvy, A. A. Kornyshev, S. Leikin, J. Phys. Chem. B 
 {\bf 106}, 13362 (2002).  
\bibitem{lyubar} A. P. Lyubartsev, L. Nordenski{\"o}ld,
  J. Phys. Chem. B {\bf 101}, 4335 (1997).
\bibitem{montoro1998} J. C. G. Montoro, J. L. F. Abascal, J. Chem. Phys. {\bf
  103}, 8273 (1995), J. Chem. Phys. {\bf 109}, 6200 (1998); 
 J. L. F. Abascal, J. C. G. Montoro,  J. Chem. Phys. {\bf 114}, 4277 (2001).
\bibitem{gulbrand1989} L. E. Guldbrand, T. R. Forester,
  R. M. Lynden-Bell, Mol. Phys. {\bf 67}, 473 (1989).
\bibitem{jaya2} B. Jayaram, K. Sharp, B. Honig, Biopolymers {\bf 28},
  975 (1989).
\bibitem{conrad1988biopol} J. Conrad, M. Troll, B. H. Zimm, Biopolymers {\bf 27},
  1711 (1988).
\bibitem{hochberg} D. Hochberg, T. W. Kephart, G. Edwards, Phys. Rev. E {\bf
  49}, 851 (1994); D. Hochberg, G. Edwards, T. W. Kephart, Phys. Rev. E
  {\bf 55}, 3765 (1997); G. Edwards, D. Hochberg, T. W. Kephart,
  Phys. Rev. E {\bf
    50}, R698 (1994).
\bibitem{zakharova} S. S. Zakharova, S. U. Egelhaaf, L. B. Bhuiyan,
  D. Bratko, J. R. C. van der Maarel, J. Chem. Phys. {\bf 111}, 10706 (1999).
\bibitem{lukatsky2002epl} D. B. Lukatsky, S. A. Safran, A. W. C. Lau,
  P. Pincus, Europhys. Lett. {\bf 58}, 785 (2002).
\bibitem{kjellander2001solvent} R. Kjellander, A. P. Lyubartsev,
S. Mar\v cella, J. Chem. Phys. {\bf 114}, 9565 (2001). 
\bibitem{messina1} R. Messina, C. Holm, K. Kremer Phys. Rev. E {\bf 64},
 021405 (2001); {\it ibid} Eur. Phys. J. E {\bf 4}, 363 (2001); R. Messina,
  C. Holm, K. Kremer, Phys. Rev. Lett. {\bf 85}, 872 (2000).
\bibitem{sponer2002} J. Sponer, J. Leszczynski, P. Hobza, Biopolymers
  {\bf 61}, 3 (2002).
\bibitem{duguid1993} J. Duguid, V. A. Bloomfield, J. Benevides,
  G. J. Thomas Jr, Biophys. J. {\bf 65}, 1916 (1993).
\bibitem{duguid1995} J. G. Duguid, V. A. Bloomfield et al,
  Biophys. J. {\bf 69}, 2623 (1995); ibid Biophys. J. {\bf 69}, 2642 (1995).
\bibitem{knoll1988book} D. A. Knoll, M. D. Fried, V. A. Bloomfield "DNA and its Drug
  Complexes" edited by R. H. Sarma and M. H. Sarma (Adenin press,
  New-York 1988, p.123)
\bibitem{rau1992biophys} D. C. Rau, V. A. Parsegian, Biophys J. {\bf 61},
  260 (1992); {\it ibid} {\bf 61}, 246 (1992). 
\bibitem{widom} J. Widom, R. L. Baldwin, J. Mol. Biol. {\bf 144}, 431 (1980).
\bibitem{plum} G. E. Plum, V. A. Bloomfield, Biopolymers {\bf 27},
  1045 (1988); ibid {\bf 29}, 13 (1990); ibid {\bf 30}, 631 (1990).
\bibitem{braunlin1986biopolymers} W. H. Braunlin, C. F. Anderson,
  M. T. Record Jr., Biopolymers {\bf 25}, 205 (1986); W. H. Braunlin,
  Q. Xu, Biopolymers {\bf 32}, 1703 (1992).
\bibitem{arscott1995biopolymers} P. G. Arscott, C. Ma, J. R. Wenner,
V. A. Bloomfield, Biopolymers {\bf 36}, 345 (1995).
  J. Phys. Chem. B {\bf 102}, 7666 (1998).
\bibitem{cohen1998book} S. S. Cohen, A Guide to Polyamines, Oxford
  University Press, New-ork, NY, 1998
\bibitem{tabor1984} C. W. Tabor, H. Tabor, Annu. Rev. Biochem. {\bf
  53}, 749 (1984).
\bibitem{marx1982} K. A. Marx, T. C. Reynolds,
  Proc. Natl. Acad. Sci. USA {\bf 79}, 6484 (1982).
\bibitem{saminathan} M. Saminathan, T. Thomas, A. Shirahata,
  C. K. S. Pillai, T. J. Thomas, Nucleic
Acids Research {\bf 30}, 3722 (2002).
\bibitem{gosule1976and1978} L. C. Gosule, J. A. Schellman, Nature {\bf
    259}, 333 (1976); J. Mol. Biol. {\bf 121}, 311 (1978).
\bibitem{wilson1} R. W. Wilson, V. A. Bloomfield, Biochemistry {\bf 18},
  2192 (1979);  R. W. Wilson, D. C. Rau, V. A. Bloomfiled,
  Biophys. J. {\bf 30}, 317 (1980). 
\bibitem{deng2000} H. Deng, V. A. Bloomfield, J. M. Benevides,
  G. J. Thomas Jr, Nucleic Acids Research {\bf 28}, 3379 (2000).
\bibitem{raspaud1999} E. Raspaud, I. Chaperon, A. Leforestier,
  F. Livolant, Biophys. J. {\bf 77}, 1547 (1999); 
E. Raspaud, M. Olvera, M. O. de la Cruz, J.-L. Sikorav, F. Livolant, 
 Biophys. J. {\bf 74}, 381 (1998).
\bibitem{anderson1982arpc} C. F. Anderson, M. T. Record Jr.,
  Anu. Rev. Phys. Chem. {\bf 33}, 191 (1982). 
\bibitem{manning1} G. S. Manning, Q. Rev. Biophys. {\bf 11}, 179
  (1978); Acc. Chem. Res. {\bf 12}, 443 (1979).
\bibitem{manning1992} J. Ray, G. S. Manning, Biopolymers {\bf 32}, 541 (1992).
\bibitem{rouzina2} I. Rouzina, V. A. Bloomfield, J. Chem. Phys. {\bf 100},
  4292 (1996).
\bibitem{korolev} N. Korolev, A. P. Lyubartsev, L. Nordenski\"{o}ld,
  A. Laaksonen, J. Mol. Biol. {\bf 38}, 907 (2001). 
\bibitem{andrey} Here we adopt an idealized DNA model where phosphates
  form ideal spirals. It is however known that in real DNAs the twist
  angle from one base pair to another depends on the base pair
  sequence. The influence of the twist angle fluctuations on the
  DNA-DNA interaction is addressed in  A. G. Cherstvy, A. A. Kornyshev,
 S. Leikin, submitted to J. Phys. Chem. B (from private communication with
  A. G. Cherstvy). 
\bibitem{bonvin2000ebj} A. M. J. J. Bonvin, Euro. Biophys J. {\bf
  29}, 57 (2000). 
\bibitem{Hartmut} H. L{\"o}wen, Progr. Colloid Polym. Sci. {\bf 110}, 12 (1998).
\bibitem{wu1999} J. Z. Wu, D. Bratko, H. W. Blanch, J. M. Prausnitz
J. Chem. Phys. {\bf 111}, 7084 (1999).
\bibitem{olmsted1989and1995} M. C. Olmsted, C. F. Anderson, M. T. Record Jr,
  PNAS {\bf 86}, 7766 (1989); M. C. Olmsted, J. P. Bond, C. F. Anderson,
  M. T. Record, Biophys. J. {\bf 68}, 634 (1995).
\bibitem{feig1999} M. Feig, B. M. Pettitt, Biophys. J. {\bf 77}, 1769 (1999).
\bibitem{allison1994} S. A. Allison, J. Chem. Phys. {\bf 98},
  12091 (1994).
\bibitem{hingerty} B. E. Hingerty, R. H. Ritchie, T. L. Ferrel, J. E. Turner,
  Biopolymers {\bf 24}, 427 (1985).
\bibitem{Tandon} S. Tandon, R. Kesavamoorthy, S. A. Asher,
  J. Chem. Phys. {\bf 109}, 6490 (1998). 
\bibitem{lekner} J. Lekner, Physica A {\bf 176}, 485 (1991); J. Lekner,
  Mol. Simul. {\bf 20}, 357 (1998). 
\bibitem{stigter1996} D. Stigter, K. A. Dill,  Biophys. J. {\bf 71}, 2067 (1996).
\bibitem{fujimoto1994bj} B. S. Fujimoto, J. M. Miller, N. S. Ribeiro,
  J. M. Schurr, Biophys. J. {\bf 67}, 304 (1994).
\bibitem{ninham1997langmuir} B. W. Ninham, V. Yaminsky, Langmuir {\bf
    13}, 3097 (1997). 
\bibitem{protein2} R. Piazza, M. Pierno, J. Phys. Condensed Matter 
  {\bf 12}, A443 (2000).  
\bibitem{pettitt} T. Cagin, J. I. Pettitt, Mol. Phys. {\bf 72}, 169 (1991);
 T. Cagin, J. I. Pettitt, J. Chem. Phys. {\bf 96}, 1333 (1992);
J. I. Pettitt, Mol. Phys {\bf 82}, 67 (1994);
G. C. Lynch, J. I. Pettitt, J. Chem. Phys. {\bf 107}, 8594 (1997);
A. Weerasinghe, J. I. Pettitt, Mol. Phys {\bf 82}, 897 (1994).
\bibitem{lo1995} C. Lo, B. J. Palmer, J. Chem. Phys, {\bf 102}, 925 (1995). 
\bibitem{allen} M. P. Allen and D. J. Tildesley, Computer simulation of
  Liquids, Oxford Science Publications, Oxford University Press,
  Oxford (1991).
\bibitem{lupkowski1991} M. Lupkowski, F. van Swol, J. Chem. Phys. {\bf
  95}, 1995 (1991). 
\bibitem{valleau1980} J. P. Valleau, L. K. Cohen, J. Chem. Phys. {\bf
  72}, 5935 (1980); J. P. Valleau, L. K. Cohen, D. N. Card,
  J. Chem. Phys. {\bf 72}, 5942 (1980).
\bibitem{korolev1999} N. Korolev,  A. P. Lyubartsev,  A. Rupprecht,
  L. Nordenski{\"o}ld, Biophys. J. {\bf 77}, 2736 (1999).
\bibitem{yau1994} Yau, Liem, Chan, J. Chem. Phys. {\bf 101}, 7918 (1994).
\bibitem{torrie1980} G. M. Torrie, J. P. Valleau, G. N. Patey,
   J. Chem. Phys. {\bf 76}, 4615 (1982).
\bibitem{mon1985} K. K. Mon, R. B. Griffiths, Phis. Rev. A {\bf 31}, 956 (1985).
\bibitem{orkoulas1994} G. Orkoulas, A. Panagiotopoulos, J. Chem. Phys.
  {\bf 101}, 1452 (1994).
\bibitem{papadopoulou1993}  A. Papadopoulou, E. D. Becker,
  M. Lupkowski, F. van Swol, J. Chem. Phys. {\bf 98}, 4897 (1993).  
\bibitem{attard1997} P. Attard, J. Chem. Phys. {\bf 107}, 3230 (1997),
  and references therein.
\bibitem{Svennson_gcmc} B. R. Svennson, C. E. Woodward, Molecular
   Physics {\bf 64}, 247 (1988).
\bibitem{jonsson_lecture} B. Jonsson, Lecture notes, Les houches,
 October (2000).
\bibitem{zipper} A. A. Kornyshev, S. Leikin, Phys. Rev. Letters {\bf
    82}, 4138 (1999);  Phys. Rev. Letters {\bf 86}, 3666 (2001).  
\bibitem{schellman1984crosslink} J. A. Schellman, N. Parthasaratary,
J. Mol. Biol. {\bf 175}, 313 (1984).
\bibitem{linse2002attraction} P. Linse, J. Phys. Cond. Matter {\bf
  14}, 13449 (2002). 
\bibitem{HansenLoewen} J. P. Hansen, H. L\"owen, Annual Rev. Phys. Chem. 
{\bf 51}, 209 (2000). 
\bibitem{hydration_attraction} This effect resembles the so called
  ``hydration attraction'' \cite{rau1992biophys}, when the entropic
  ``push'' from the outside area is generated by the water particles. A
  similar effect, called ``a counterion evaporation''(see C. Fleck,
  H. H. von Gr\"unberg, Phys. Rev. E {\bf 63}, 061804 (2001)) exists
  between the oppositely charged rod and surface.
\bibitem{PuseyLH}   P. N. Pusey, in ``Liquids, Freezing and
the Glass Transition",
edited by J. P. Hansen, D. Levesque and J. Zinn-Justin (North Holland, Amsterdam,
1991).
\bibitem{allahprotein} E. Allahyarov, H. L{\"o}wen, J.-P. Hansen,
  A. A. Louis, Phys. Rev. E {\bf 67}, 051404 (2003).
\bibitem{moon2000salteffect} Y. U. Moon, C. O. Anderson, H. W. Blanch,
  J. M. Prausnitz, Fluid Phase Equilibria {\bf 168}, 229 (2000).
\bibitem{manning1978salt} G. S. Manning, Q. Rev. Biophys. {\bf 11},
179-246 (1978).
\bibitem{wu1998} J. Z. Wu, D. Bratko, H. W. Blanch, J. M. Prausnitz,
  Proc. Natl. Acad. Sci. Usa {\bf 95}, 15169 (1998). 
\bibitem{belloni1995} L. Belloni et all, J. Chem. Phys. {\bf 103},  5781 (1995).  
\bibitem{sottas1999bj} P. E. Sottas, E. Larquet, A. Stasiak,
  J. Dubochet, Biophys. J. {\bf 77}, 1858 (1999).
\bibitem{nguen_reentrant} T. T. Nguyen, I. Rouzina, B. I. Shklovskii,
  J. Chem. Phys. {\bf 112}, 2562 (2000);
T. T. Nguyen, B. I. Shklovskii, J. Chem. Phys, {\bf 115}, 7298 (2001).
\bibitem{gast1983} A. P. Gast, W. B. Russel, C. K. Hall,
  J. Colloid. Interface Sci. {\bf 96}, 251 (1983); ibid {\bf 109}, 161
  (1986). 
\bibitem{lekker1992}  H. N. W. Lekkerkerker, W. C. K. Poon,
  P. N. Pusey, A. Stroobants, P. B. Warren, Europhys. Lett. {\bf 20}, 559 (1992).
\bibitem{potemkin2002} I. Potemkin, A. R. Khohklov et al, Phys, Rev, E
  {\bf 66}, 011802 (2002).
\bibitem{ha1997nonadditivity} B.-Y. Ha, A. J. Liu, Phys. Rev. Letters
  {\bf 79}, 1289 (1997).
\bibitem{knott2001PBattraction} M. Knott, I. J. Ford, Phys. Rev. E {\bf
  63}, 031403 (2001). 
\bibitem{warren2000jcp} P. B. Warren, J. Chem. Phys. {\bf 112},
  4683 (2000).
\bibitem{dan1996bj} N. Dan, Biophys. J. {\bf 71}, 1267 (1996).
\bibitem{chirality} All helices are chiral, i.e., the mirror image of a right-handed
  helix is left handed and vice versa.
\bibitem{strzelecka1987and1988} T. E. Strzelecka, R. L. Rill,
J. Am. Chem. Soc. {\bf 109}, 4513 (1987).
\bibitem{livolant} F. Livolant, Physica A {\bf 176},
  117 (1991); A. Leforestier, F. Livolant, Biophys. J. {\bf 65}, 56
  (1993); F. Livolant, A. Leforestier, Mol. Cryst. Liq. Cryst. {\bf
    215}, 47 (1992); D. Durand, J. Doucet, F. Livolant, J. Phys. II
  {\bf 2}, 1769 (1992). 
\bibitem{simulationswithmolecularwater} L. R. Pratt, G. Hummer,
  A. E. Garsia, Biophys. Chem. {\bf 51}, 147 (1994); E. Guardia,
  R. Rey, J. A. Padro, J. Chem. Phys. {\bf 95}, 2823 (1991).
\bibitem{lyubartsev1995pre} A. P. Lyubartsev, A. Laaksonen,
  Phys. Rev. E {\bf 52}, 3730 (1995).
\bibitem{belloni1997} L. Belloni, O. Spalla, J. Chem. Phys. {\bf 107}, 465 (1997).
\bibitem{kusalik1988} P. G. Kusalik, G. N. Patey,  J. Chem. Phys. {\bf
  88}, 7715 (1988). 
\bibitem{allahyarovsolventeffect} E. Allahyarov, H. L{\"o}wen,
Phys. Rev. E {\bf 63}, 041403 (2001).
\bibitem{pettitt1986jcp} B. M. Pettitt, P. J. Rossky
  J. Chem. Phys. {\bf 84}, 5836 (1986). 
\bibitem{LIE} F. Otto, G. N. Patey, Phys. Rev. E {\bf 60}, 4416 (1999); 
J. Chem. Phys. {\bf 112}, 8939 (2000).
\bibitem{chiu2000jmb} T. K. Chiu, R. E. Dickerson, J. Mol. Biol. {\bf
  301}, 915 (2000).
\bibitem{korolev2001biopolymers} N. Korolev, A. P. Lyubartsev,
  A. Rupprecht, L. Nordenski{\"o}ld, Biopolymers {\bf 58}, 268 (2001).
\bibitem{korolev2002biophys} N. Korolev, A. P. Lyubartsev,
A. Laaksonen, L. Nordenski\"old, Biophys. J. {\bf 82}, 2860 (2002).
\bibitem{young2} M. A. Young, D. L. Beveridge,
  J. Mol. Biol. {\bf 281}, 675 (1998).
\bibitem{linse1986} P. Linse, J. Phys. Chem. {\bf 90}, 6821 (1986).
\bibitem{saenger1987} W. Saenger, Ann. Rev. Biophys. Chem. {\bf 16} 93
  (1987). 
\bibitem{jayaram1998jpc} M. A. Young, B. Jayaram, D. L. Beveridge,
\bibitem{maarel1999} J. R. C. van der Maarel, Biophysical J. {\bf 76}, 2673
  (1999).
\bibitem{mazur} J. Mazur, R. L. Jernigan, Biopolymers {\bf 31}, 1615 (1991).
\bibitem{petsev2000bj} D. N. Petsev, B. R. Thomas, S. T. Yau,
  P. G. Vekilov, Biophys. J. {\bf 78}, 2060 (2000).
\bibitem{luka1} A. V. Lukashin, D. B. Beglov, M. D. Frank-Kamenetskii,
  J. Biomolecular Structure and Dynamics {\bf 9}, 517 (1991).
\bibitem{jayaram1996epsilonR} B. Jayaram, A. Das, N. Aneja,
  J. Mol. Structure {\bf 361}, 249 (1996).
\bibitem{hydration_forces}S. Leikin, D. C. Rau, V. A. Parsegian,
  Phys. Rev. A {\bf
  44}, 5272 (1991); R. Podgornik, D. Rau, V. A. Parsegian,
  Biophys J. {\bf 66}, 962 (1994); R. P. Rand, N. Fuller, V. A. Parsegian, D. C. Rau,
  Biochemistry {\bf 27}, 7711 (1988); C. V. Valdeavella, H. D. Blatt, L. Yang,
  B. M. Pettitt, Biopolymers {\bf 50}, 133 (1999); P. Mariani, L. Saturni,  Biophysical  J. {\bf 70}, 2867 (1996).
\bibitem{petsev2000} D. N. Petsev, P. G. Vekilov, Phys. Rev. Letters
  {\bf 84}, 1339  (2000).
\bibitem{ibragimova1998} G. T. Ibragimova, R. C. Wade,
  Biophys. J. {\bf 74}, 2906 (1998).
\bibitem{liu2000jcp} H. Liu, J. Gapinski, L. Skibinska, A. Patkowski,
  R. Pecora, J. Chem. Phys. {\bf 113}, 6001 (2000).
\bibitem{williams} L. D. Williams, L. J. Maher III,
  Annu. Rev. Biophys. Biomol. Struct. {\bf 29}, 497 (2000). 
\bibitem{lu2002} Y. Lu, B. Weers, N. C. Stellwagen, Biopolymers {\bf
 61}, 261 (2002).  
\bibitem{baumann1997} C. G. Baumann, S. B. Smith, V. A. Bloomfield,
  C. Bustamante, Proc. Natl. Acad. Sci. Usa {\bf 94}, 6185 (1997).
\end{references}
\end{document}